\documentclass[
reprint,
bibnotes,
amsmath,amssymb,
aps,
prd
]{revtex4-1}

\usepackage{amssymb,amsmath,mathtools}
\usepackage{amsthm}
\usepackage{titletoc}
\usepackage{graphicx}
\usepackage{dcolumn}
\usepackage{enumitem}
\usepackage{bm}
\usepackage{float}
\usepackage[many]{tcolorbox}
\usepackage{shuffle}
\usepackage{xcolor}
\usepackage{physics}
\usepackage{tabularx}
\usepackage{diagbox}
\usepackage{subcaption}
\usepackage{multirow}
\usepackage{caption}
\usepackage{tikz}
\usepackage{tikz-feynman}
\usepackage{hyperref}
\usepackage{comment}

\usetikzlibrary{decorations.pathmorphing}
\usetikzlibrary{arrows}
\usepackage{circuitikz}
\usetikzlibrary{arrows.meta}
\usetikzlibrary{shapes.misc}
\usetikzlibrary{positioning}
\usetikzlibrary{decorations.markings}
\usetikzlibrary{intersections}
\usetikzlibrary{calc}
\usepackage{verbatim}
\tikzset{
    vector/.style={
        decoration={snake, aspect=0.75, mirror, segment length=2mm},
        decorate
    },
    photon/.style={decorate, decoration={snake, amplitude=1pt, segment length=4pt}}
}

\tikzset{
  cross/.style={
    path picture={
      \draw
        (path picture bounding box.south west) --
        (path picture bounding box.north east)
        (path picture bounding box.north west) --
        (path picture bounding box.south east);
    }
  }
}

\usepackage{CJK}
\definecolor{Maroon}{RGB}{128, 0, 0}
\definecolor{Green}{rgb}{0.0, 0.5, 0.0}
\definecolor{Purple}{rgb}{0.5, 0.0, 0.5}
\definecolor{Blue}{rgb}{0.0, 0.0, 0.6}

\begin{document}

\begin{CJK*}{UTF8}{}
\CJKfamily{gbsn}

\title{Generating the fermion mass hierarchy at the TeV scale
}

\author{Nima Arkani-Hamed$^{1}$}
\email{arkani@ias.edu}
\author{Carolina Figueiredo$^{2}$}
\email{cfigueiredo@princeton.edu}
\author{Lawrence J. Hall$^{3,4}$}
\email{ljh@berkeley.edu}
\author{Claudio Andrea Manzari$^{1}$}
\email{manzari@ias.edu}

\affiliation{$^{1}$School of Natural Sciences, Institute for Advanced Study, Princeton, NJ, 08540, USA \\
$^{2}$Jadwin Hall, Princeton University, Princeton, NJ, 08540, USA
\\
$^{3}$Leinweber Institute for Theoretical Physics, Department of Physics, University of California, Berkeley, CA 94720, USA
\\
$^{4}$Theoretical Physics Group, Lawrence Berkeley National Laboratory, Berkeley, CA 94720,
USA}

\begin{abstract}
We propose a class of theories to generate quark and lepton mass matrices where the scale of new physics is at the TeV scale, without inducing the large flavor and CP violating processes that are often thought to  relegate the origin of flavor to energies above $\sim 100$ TeV. The models have new vector-like leptons and quarks, with mass mixings to each other and Yukawa couplings to light Standard Model fields encoded in ``chains" reminiscent of dimensional deconstruction.  Locality in the chains both generates the hierarchical Standard Model Yukawa matrices, and ensures that CP and flavor violating effects are small, even with the vector-like particles at the TeV scale. A simple extension also generates neutrino masses, whose tiny size is parametrically related to the square of the electron Yukawa coupling. We outline the essential features of these models, explain how fermion mass hierarchies and mixing angles emerge, and explore their phenomenological implications. This framework can be tested both in the final run of the LHC as well as at possible future colliders operating at the 10 TeV scale, and we identify some of the distinctive experimental signatures associated with the production and decay of the new vector-like fermions. 
\end{abstract}
\maketitle
\end{CJK*}

\section{Introduction}

The flavor structure of elementary particles remains one of the most persistent open problems in fundamental physics. The mystery encompasses several intertwined questions: why do fermions come in three generations? What is the origin of neutrino masses? And what determines the hierarchies observed among fermion masses and mixing angles? 

Within the Standard Model (SM), fermion masses arise from Yukawa interactions with the Higgs field. While successful, this mechanism leaves the values of these couplings, spanning from ${\cal O}(1)$ for the top quark to ${\cal O}(10^{-6})$ for the electron, entirely unexplained. Anthropic arguments offer limited insight, potentially constraining light fermion masses relevant for the properties of stable matter, but failing to account for the overall pattern of masses and mixings~\cite{Hogan:1999wh,Haba:2000be,Donoghue:2005cf,Hall:2007ja, Donoghue:2007zz, Jaffe:2008gd, Jenkins:2009ux, Berengut:2013nh, HossainAli:2012eof}. Conventional theoretical approaches, such as flavor symmetries~\cite{Pakvasa:1977in,Froggatt:1978nt,Chivukula:1987py,Leurer:1992wg,Barbieri:1995uv,Barbieri:1996ww,Barbieri:1997tu,Barbieri:1997tg,Barbieri:1998em,Barbieri:1998qs,Altarelli:2010gt,Feruglio:2017spp,Barbieri:2019zdz,Novichkov:2021evw,Greljo:2025mwj,Banks:2026amr} or extra spatial dimensions~\cite{Arkani-Hamed:1999ylh,Arkani-Hamed:1999pwe,Mirabelli:1999ks,Delgado:1999sv,Gherghetta:2000qt,Bando:2000it,Huber:2000ie,Agashe:2004cp,Csaki:2008zd,Glioti:2024hye,Agashe:2025tge}, often imply that the dynamics responsible for flavor reside at energy scales well above the electroweak scale. This expectation is driven by strong constraints on flavor-changing neutral currents (FCNCs) and CP violation, which generally force generic new flavor dynamics to scales exceeding ${\cal O}(10^2-10^4)$~TeV. Consequently, the prospects for directly detecting the origin of flavor at the LHC or future colliders are typically bleak.

Counter to this long-standing lore that flavor can only be generated at inaccessibly high energies scales, in this letter we propose a very simple class of theories for flavor where the scale of new physics can be as low as the TeV scale; some other proposals for low-scale theories of flavor can be found in \cite{Leurer:1992wg,Arkani-Hamed:1999pwe,Greljo:2025mwj}.  Our theories have a very minimal content--augmenting the SM  with a set of vector-like (VL) quarks and leptons.  The theory is governed by a flavor symmetry $G_F$ that forbids Yukawa couplings between two light SM fermions. The observed SM Yukawa couplings are instead generated by soft $G_F$-breaking mass terms, of order $\mu$, that mix heavy and light states. The pattern of soft masses as well as Yukawa couplings between the heavy fields and the light quarks and leptons are efficiently encoded in a ``chain diagram" familiar from dimensional deconstruction~\cite{Arkani-Hamed:2001kyx}; the soft masses allow the fields to ``hop" along the chain. The resulting light fermion masses and mixings arise from the interplay of the allowed mass terms and light-heavy Yukawa couplings, $\lambda$.\\
\\
\indent While this framework allows for diverse structures, such as $ G_F = U(1)^{15}$ with three full VL generations or $G_F=U(2)$ with doublet VL generations, we restrict our attention in this work to a specific, minimal realization. We consider models where the VL fermions are $SU(2)$ gauge singlets ($U^c, D^c, E^c$) and the flavor symmetry is Abelian, restricting the allowed interactions to a few light-heavy terms (e.g., $\lambda \, q U^c h^\dagger$) and specific heavy-heavy mass mixings (e.g., $\mu \, U_1 U^c_2)$. Despite the low mass scale, the simple site-link structure of the resulting chains renders CP violation benign and naturally suppresses flavor-violating effects in both quark and lepton sectors. We find that a mass scale $M\sim$ TeV is consistent with $\lambda, \mu/M \approx 0.1-0.2$. There are many simple field theoretic mechanisms for generating these couplings in the deep UV, and indeed it is not even unreasonable to think this modest small parameters already emerge in the low-energy theory beneath the Planck/string scales. Crucially, this ensures that the heavy fermions are within reach of the final run of LHC and future colliders operating at energies of order $10$~TeV, offering a direct experimental window into the origin of flavor hierarchies.
\begin{figure*}[t]
    \centering
    \includegraphics[width=\linewidth]{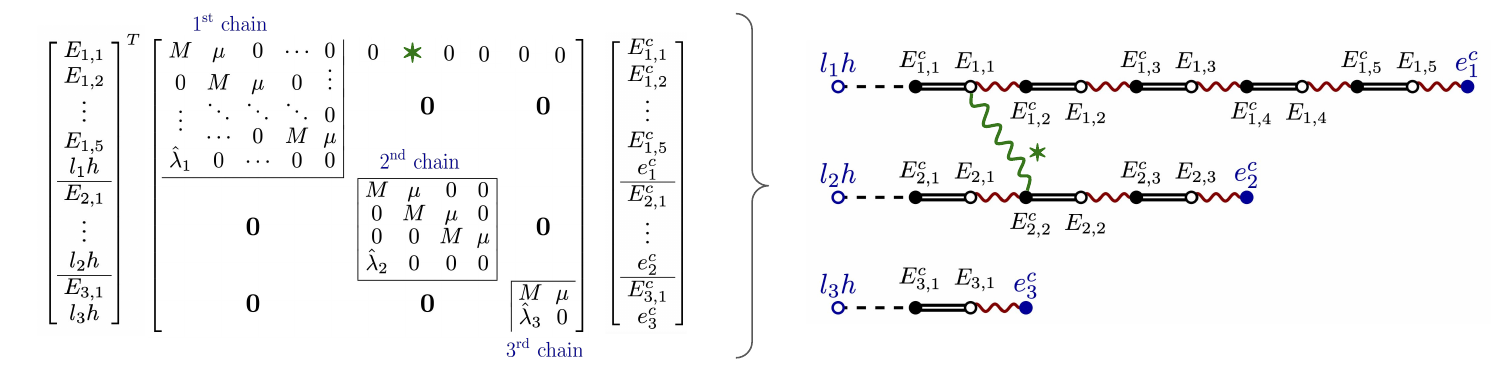}
    \caption{\textbf{(Left)} Mass and Yukawa matrices for VL and light fermions. \textbf{(Right)} Graphical representation of interactions via chains. The links of the chains are of three types: dashed lines represent Yukawa couplings, $\lambda_i l_i E^c_{i,1} h$, the double solid lines represent vector-like masses, $M_{I} E_{I}^c E_{I}$, and the wiggly lines denote soft breaking off-diagonal mass mixing, $\mu^E_{IJ} E_{I} E^c_{J}$. In green, we represent a link between chains that generates a non-zero off-diagonal entry in the mass matrix. }
    \label{fig:LeptonChains}
\end{figure*}

\section{Chain Models}

We begin by illustrating our ``chain models". An abelian flavor symmetry allows for the following Lagrangian in the lepton sector,
\begin{equation}
\begin{aligned}
  {\cal L}_F = \hat{\lambda}_{iI}^E \, \ell_i E^c_I h + M_I^E \, E_I E^c_I\,,
\end{aligned}
\label{eq:LEGFinv}
\end{equation}
where $h$ is the SM Higgs field, $i=1,2,3$, and $I$ runs over however many VL-states there are. Although the $e_i$ and $E_I$ states mix, the SM states remain massless until a set of soft flavor symmetry breaking mass terms are added
\begin{equation}
\begin{aligned}
  {\cal L}_{\text{soft}} = E_I \, ( \mu_{IJ}^E \, E^c_J + \mu_{Ij}^E \, e^c_j).
\end{aligned}
\end{equation} 
Since $I\neq J$ we call the $\mu$ parameters ``hopping terms"; they are a key origin of flavor changes in the theory. The $U$, $D$ sectors are very similar to the $E$ sector, with $(\ell, E, E^c) \rightarrow (q, U, U^c), (q, D, D^c)$. 
Crucially, the flavor symmetry results in the $\hat{\lambda}$ matrices having many zero entries. Similarly, a very restricted choice of symmetry breaking parameters is chosen so that the $\mu$ matrices are also very \textit{sparse}. Each non-zero entry of the SM Yukawa matrices is then generated by a \textit{chain} diagram. The SM doublet field $\ell_i$ or $q_i$ first connects to a VL fermion via $\hat{\lambda}_{iI}$; there is then a sequence of mixings between successive heavy VL states via a product of entries of $\mu_{KL}$ to take the VL fermion of type I to one of type J, and finally the chain ends with a $\mu_{Jj}$ hop connecting to $e^c_j, u^c_j, d^c_j$ (see figure \ref{fig:LeptonChains} for an example). If the non-zero entries of the $\hat{\lambda}_{iI}$ matrices are taken to be $\hat{\lambda}$, and the size of the dimensionless soft symmetry breaking parameters $\mu/M$, are taken to be $\epsilon$, then the resulting SM Yukawa coupling is 
\begin{equation}
\begin{aligned}
  y_{ij} \sim \hat{\lambda} \, \epsilon^{n_{ij}}
\end{aligned}
\end{equation}
where there are $n_{ij}$ different VL fermions in the chain that generates $y_{ij}$.
We choose the chains to be long enough that there are no very small numbers in the theory, for example $\hat{\lambda} \sim \epsilon \sim 0.1-0.2$. Thus, the observed flavor hierarchies of the SM are generated by the diagonalization of the full fermion mass matrices, rotating between flavor and mass eigenstate bases. We illustrate this first with the minimal model for the lepton sector and then by a simple model for the quark sector.
\\

\noindent {\bf Minimal chain model for lepton flavor.}--- The minimal model in the lepton sector generates only three entries in the SM Yukawa matrix $y^E$, which must have a non-zero determinant. We choose to order the $\ell_i$ and $e^c_i$ states so that $y^E$ is diagonal. The three chains of figure \ref{fig:LeptonChains} arise because the flavor symmetry allows only three non-zero entries in $\hat{\lambda}^E_{iI}$, 
yielding a diagonal SM charged lepton Yukawa matrix: 
\begin{equation}
    y^E_{ij} \; \ell_i e^c_j h , \, \, \, \text{with} \, \, \, y_{e,\mu,\tau} = \hat{\lambda}_1 \epsilon_E^6, \hat{\lambda}_2 \epsilon_E^4, \hat{\lambda}_3 \epsilon_E^2.
    \label{eq:EYukawas}
\end{equation}
We have chosen the lengths of each chain so that the observed values for $y_{e,\mu,\tau}$ result by choosing $\hat{\lambda}_i, \epsilon_E \sim 0.1-0.2$. The chains of figure \ref{fig:LeptonChains} generate the leading order contributions to the electron, muon and tau masses from diagonalization of the lepton mass matrix of the full theory, shown on the left of figure \ref{fig:LeptonChains}. The $E_I, E^c_I$ states are ordered so that hopping along the chains is nearest neighbor. Provided there are no links between the chains (we exclude the green link in figure \ref{fig:LeptonChains}), there are no flavor-violating processes, and since there are also no closed loops in the chains, we can remove the phases from all couplings/masses (links) by rephasing the fields (sites). Therefore, we also don't generate any electric dipole moments (EDMs) at low-energies. For this simple reason, this minimal model evades all the flavor constraints and the only bound on $M^E_I \sim M_E$ comes from the shift of the $Z$-coupling (see section \ref{sec:Constraints}).

Representing the mass-matrix via chains lets us immediately read off the effective low-energy Yukawas that we obtain after integrating out the $E/E^c$'s. Additionally, we can similarly read off the effective light-heavy Yukawas between $\ell_i$ and $E^c_{I}$ which are produced when we go to the heavy mass-eigenstate -- this is the non-zero entries of the Yukawa matrix in mass-eigenstate basis, which we denote by $\lambda_{i I}$. Due to the hierarchical structure of the mass-matrix, each mass eigenstate will be mostly one of the states in the chain basis, and for a given $E^c_{I}$ we can read off the effective Yukawa by drawing the appropriate diagram and evaluating it near the on-shell locus, $p^2 = - M^2_{I}$. For example, for the state which is mostly $E^c_{1,2}$ in the first chain in figure \ref{fig:LeptonChains} we get the effective Yukawa
\begin{equation}
\begin{gathered}
    \begin{tikzpicture}[line width=0.9,scale=0.35,baseline={([yshift=0.0ex]current bounding box.center)},
  fermion/.style={
    line width=0.9pt,
    postaction={
      decorate,
      decoration={
        markings,
        mark=at position 0.6 with {
          \arrow{Latex[length=5pt,width=5pt]}
        }
      }
    }
  },
  antifermion/.style={
    line width=0.9pt,
    postaction={
      decorate,
      decoration={
        markings,
        mark=at position 0.6 with {
          \arrow{Latex[length=5pt,width=5pt, reversed]}
        }
      }
    }
  },
  scalar/.style={
    dashed,
    line width=0.9pt,
    postaction={
      decorate,
      decoration={
        markings,
        mark=at position 0.5 with {
          \arrow{Latex[length=5pt,width=5pt]}
        }
      }
    }
  },
  antiscalar/.style={
    dashed,
    line width=0.9pt,
    postaction={
      decorate,
      decoration={
        markings,
        mark=at position 0.5 with {
          \arrow{Latex[length=5pt,width=5pt, reversed]}
        }
      }
    }
  },
  photon/.style={
    decorate,
    decoration={snake, amplitude=1.2pt, segment length=4pt},
    line width=0.9pt
  }
]

\coordinate (v0)  at (0,4);
\coordinate (v1)  at (2,2);
\coordinate (v2)  at (0,0);
\coordinate (v10) at (4,2);
\coordinate (v11) at (6,2);
\coordinate (v12) at (6,0);
\coordinate (v13) at (8,2);
\coordinate (v14) at (10,2);
\coordinate (v15) at (12,4);
\coordinate (v16) at (12,0);

\coordinate (Ec) at ($(v10)!0.5!(v1)$);
\node[color=black,scale=0.8,xshift=-2,yshift=9] at (Ec) {$E^c_{1,1}$};

\coordinate (E) at ($(v10)!0.5!(v11)$);
\node[color=black,scale=0.8,xshift=2,yshift=9] at (E) {$E_{1,1}$};

\coordinate (Eb) at ($(v13)!0.5!(v11)$);
\node[color=black,scale=0.8,xshift=1,yshift=9] at (Eb) {$E^c_{1,2}$};

\draw[fermion]     (v0)  -- (v1);
\draw[scalar]      (v2)  -- (v1);
\draw[fermion]     (v10) -- (v1);

\node[cross, minimum size=2pt,scale=0.8] at (v10) {};

\draw[fermion]     (v10) -- (v11);

\fill (v11)  circle [radius=0.15];
\node[color=black,scale=0.8,xshift=0,yshift=-9] at (v11) {$\mu$};
\draw[fermion] (v13) -- (v11);

\coordinate (li) at ($(v0)!0.5!(v1)$);
\coordinate (h) at ($(v2)!0.5!(v1)$);

\coordinate (lib) at ($(v15)!0.5!(v14)$);
\coordinate (hb) at ($(v16)!0.5!(v14)$);

\node[color=black,scale=0.9,xshift=5,yshift=9] at (li) {$\ell_{1}$};
\node[color=black,scale=0.9,xshift=5,yshift=-9] at (h) {$h$};

\end{tikzpicture} 
\end{gathered} = \hat{\lambda}_1  \frac{\mu}{M_{1,1}} \times \left(\frac{M_{1,1}^2}{M_{1,1}^2 - M_{1,2}^2}\right)  ,
\end{equation}
If the heavy masses $M$ are generic, then the factor in brackets is $\mathcal{O}(1)$, and we have an overall suppression of the coupling by a factor $(\mu/M) \sim \epsilon$. 
Similarly as we go further down the chain we have that the mass eigenstate that is mostly $E^c_{i,j}$ has an effective Yukawa coupling $ \sim \hat{\lambda}_i \epsilon^{j-1}$ to $\ell_i$ -- and so the size of the Yukawa for the heavy-mass eigenstates is given by the length of the path in the chain from the heavy field to the light field at the end. Thus, if the heavy masses are comparable but generic, we see that their Yukawa couplings to light fields have a hierarchy precisely as given by the chain. This conclusion differs in the interesting limit where all the massive particles are nearly degenerate, with mass splittings $\Delta M^2/M^2 \sim \epsilon$. As we approach this limit the $\epsilon$ suppression from each insertion disappear, and we are instead doing degenerate perturbation theory. In this case {\it all} the VL particles have comparable Yukawa couplings to the light fields, suppressed only by the overall factor of $\epsilon$ in the coupling at the end of the chain. In Sec.\ref{sec:Neutrinos} of SupM we discuss a simple extension of this chain model which gives rise to small neutrino masses, parametrically of order $y_e^2 v^2/M \sim 0.1 \, \mbox{eV} (\mbox{TeV}/M)$, and order unity mixing angles.
\\

\noindent {\bf A simple chain model for quark flavor.}--- In the quark sector, we manifestly need a more interesting set of chains/links as we need to both correctly reproduce the quark masses as well as the CKM matrix, which has a CP violating phase. We will now provide an example achieving this. We choose to do the mixing fully in the down sector as this leads to the most stringent experimental constraints. Even so, we will see that the new VL fermions can be as light as a TeV. Since all the mixing is coming from the down sector, the up sector Yukawa matrix is purely diagonal, and so we can generate it in the same way we did for the leptons -- using a separate chain per generation, where instead of hopping $E/E^c$, we hop $U/U^c$ -- and where the size of the chain is determined by the size of the up-, charm- and top-quark Yukawa couplings. The only difference is that since the latter is already of $\mathcal{O}(1)$, we don't need any chain for the third generation, and so we directly have the light-light Yukawa, $y_3 q_3 u^c_3 h^\dagger$.  

As for the down-sector chains, these should be such that the low-energy Yukawa matrix yields the observed down quark masses as well as the full CKM matrix. For simplicity, we ask for the low-energy $3\times 3$ matrix to have exactly $6$ real non-zero entries and one phase, corresponding to the three masses and the 4 parameters of the CKM matrix (3 real and 1 phase). As mentioned earlier, we can keep track of the rephase invariants in our chains by looking at the number of closed loops. So a chain model with the desired number of parameters is one where there are exactly \textit{six} paths connecting $q$'s and $d^c$'s as well as  \textit{one} loop~\footnote{of course one needs to ensure that the phase in the Yukawa matrix survives in the respective CKM.}. An example of such a model is the following:
\begin{equation}
\begin{aligned}
    &\begin{gathered}
     \begin{tikzpicture}[line width=1,scale=0.22,baseline={([yshift=0.0ex]current bounding box.center)}]
\node[circle, fill=Blue, draw= Blue,  inner sep=0pt, minimum size=3.5pt] (0) at (0, 8) {};
\node[color=Blue,scale=0.9,xshift=9,yshift=-8] at (0) {$d^c_2$}; 
\node[circle,  fill=white, draw=black,  inner sep=0pt, minimum size=3.5pt] (1) at (0, 6) {};
\node[circle,  fill=black, draw=black,  inner sep=0pt, minimum size=3.5pt] (2) at (0, 4) {};
\node[circle,  fill=white, draw=black,  inner sep=0pt, minimum size=3.5pt] (3) at (0, 2) {};
\node[circle,  fill=black, draw=black,  inner sep=0pt, minimum size=3.5pt] (4) at (0, 0) {};

\node[circle,  fill=white, draw=black,  inner sep=0pt, minimum size=3.5pt] (5) at (2, 8) {};
\node[circle,  fill=white, draw=black,  inner sep=0pt, minimum size=3.5pt] (6) at (6, 8) {};
\node[circle,  fill=black, draw=black,  inner sep=0pt, minimum size=3.5pt] (7) at (4, 8) {};
\node[circle,  fill=black, draw=black,  inner sep=0pt, minimum size=3.5pt] (8) at (8, 8) {};
\node[circle,  fill=white, draw=black,  inner sep=0pt, minimum size=3.5pt] (9) at (10, 8) {};
\node[circle,  fill=black, draw=black,  inner sep=0pt, minimum size=3.5pt] (10) at (12, 8) {};
\node[circle, fill=white, draw=Blue, thick, inner sep=0pt, minimum size=3.5pt] (11) at (14, 8) {};
\node[color=Blue,scale=0.9,xshift=-4,yshift=-8] at (11) {$q_2h$}; 

\node[circle,  fill=white, draw=black,  inner sep=0pt, minimum size=3.5pt] (12) at (2, 0) {};
\node[circle,  fill=black, draw=black,  inner sep=0pt, minimum size=3.5pt] (13) at (4, 0) {};
\node[circle,  fill=white, draw=black,  inner sep=0pt, minimum size=3.5pt] (14) at (6, 0) {};
\node[circle,  fill=black, draw=black,  inner sep=0pt, minimum size=3.5pt] (15) at (8, 0) {};
\node[circle, fill=white, draw=Blue, thick, inner sep=0pt, minimum size=3.5pt] (16) at (10, 0) {};
\node[color=Blue,scale=0.9,xshift=0,yshift=9] at (16) {$q_1h$};

\node[circle,  fill=black, draw=black,  inner sep=0pt, minimum size=3.5pt] (17) at (12, 0) {};
\node[circle,  fill=white, draw=black,  inner sep=0pt, minimum size=3.5pt] (18) at (14, 0) {};
\node[circle,  fill=black, draw=black,  inner sep=0pt, minimum size=3.5pt] (19) at (16, 0) {};
\node[circle,  fill=white, draw=black,  inner sep=0pt, minimum size=3.5pt] (20) at (18, 0) {};
\node[circle,  fill=black, draw=black,  inner sep=0pt, minimum size=3.5pt] (21) at (20, 0) {};
\node[circle,  fill=white, draw=black,  inner sep=0pt, minimum size=3.5pt] (22) at (22, 0) {};
\node[circle,  fill=black, draw=black,  inner sep=0pt, minimum size=3.5pt] (23) at (24, 0) {};
\node[circle,  fill=white, draw=black,  inner sep=0pt, minimum size=3.5pt] (24) at (26, 0) {};
\node[circle, fill=Blue, draw= Blue,,  inner sep=0pt, minimum size=3.5pt] (25) at (28, 0) {};
\node[color=Blue,scale=0.9,xshift=2,yshift=9] at (25) {$d^c_3$}; 

\node[circle,  fill=white, draw=black,  inner sep=0pt, minimum size=3.5pt] (26) at (26, 2) {};
\node[circle,  fill=black, draw=black,  inner sep=0pt, minimum size=3.5pt] (27) at (24, 4) {};
\node[circle,  fill=white, draw=black,  inner sep=0pt, minimum size=3.5pt] (28) at (22, 4) {};
\node[circle,  fill=black, draw=black,  inner sep=0pt, minimum size=3.5pt] (29) at (20, 4) {};
\node[circle,  fill=white, draw=black,  inner sep=0pt, minimum size=3.5pt] (30) at (18, 4) {};
\node[circle,  fill=black, draw=black, inner sep=0pt, minimum size=3.5pt] (31) at (16, 6) {};

\node[circle,  fill=black, draw=black,  inner sep=0pt, minimum size=3.5pt] (32) at (16, 8) {};
\node[circle,  fill=white, draw=black,  inner sep=0pt, minimum size=3.5pt] (33) at (18, 8) {};
\node[circle,  fill=black, draw=black,  inner sep=0pt, minimum size=3.5pt] (34) at (20, 8) {};
\node[circle,  fill=white, draw=black,  inner sep=0pt, minimum size=3.5pt] (35) at (22, 8) {};
\node[circle,  fill=black, draw=black,  inner sep=0pt, minimum size=3.5pt] (36) at (24, 8) {};
\node[circle,  fill=white, draw=black,  inner sep=0pt, minimum size=3.5pt] (37) at (26, 8) {};
\node[circle,  fill=black, draw=black,  inner sep=0pt, minimum size=3.5pt] (38) at (28, 8) {};
\node[circle,  fill=white, draw=black,  inner sep=0pt, minimum size=3.5pt] (39) at (30, 8) {};
\node[circle, fill=Blue, draw= Blue,,  inner sep=0pt, minimum size=3.5pt] (40) at (32, 8) {};
\node[color=Blue,scale=0.9,xshift=0,yshift=-10] at (40) {$d^c_1$}; 

\node[circle,  fill=white, draw=black,  inner sep=0pt, minimum size=3.5pt] (41) at (30, 0) {};
\node[circle,  fill=black, draw=black,  inner sep=0pt, minimum size=3.5pt] (42) at (32, 0) {};
\node[circle, fill=white, draw=Blue, thick, inner sep=0pt, minimum size=3.5pt] (43) at (32, 2) {};
\node[color=Blue,scale=0.9,xshift=0,yshift=9] at (43) {$q_3h$}; 

\draw[color=Maroon,photon] (0) -- (1);
\draw[color=Maroon,photon] (0) -- (5);
\draw[double]  (1) -- (2);
\draw[double]  (3) -- (4);
\draw[double]  (12) -- (13);
\draw[double]  (14) -- (15);
\draw[double]  (5) -- (7);
\draw[double]  (6) -- (8);
\draw[double]  (9) -- (10);

\draw[color=Maroon,photon] (8) -- (9);
\draw[color=Maroon,photon] (7) -- (6);
\draw[color=Maroon,photon] (2) -- (3);
\draw[color=Maroon,photon] (4) -- (12);
\draw[color=Maroon,photon] (13) -- (14);

\draw[dash pattern=on 2pt off 1.5pt] (15) -- (16);
\draw[dash pattern=on 2pt off 1.5pt] (10) -- (11);
\draw[dash pattern=on 2pt off 1.5pt] (16) -- (17);
\draw[dash pattern=on 2pt off 1.5pt] (11) -- (32);

\draw[double]  (17) -- (18);
\draw[color=Maroon,photon] (18) -- (19);
\draw[double]  (19) -- (20);
\draw[double]  (21) -- (22);
\draw[double]  (23) -- (24);

\draw[color=Maroon,photon] (20) -- (21);
\draw[color=Maroon,photon] (22) -- (23);
\draw[color=Maroon,photon] (24) -- (25);
\draw[color=Maroon,photon] (26) -- (25);

\draw[double]  (27) -- (26);
\draw[double]  (29) -- (28);
\draw[double]  (31) -- (30);

\draw[color=Maroon,photon] (30) -- (29);
\draw[color=Maroon,photon] (28) -- (27);

\draw[dash pattern=on 2pt off 1.5pt] (11) -- (31);

\draw[double]  (32) -- (33);
\draw[double]  (34) -- (35);
\draw[double]  (36) -- (37);
\draw[double]  (38) -- (39);

\draw[color=Maroon,photon] (37) -- (38);
\draw[color=Maroon,photon] (35) -- (36);
\draw[color=Maroon,photon] (33) -- (34);
\draw[color=Maroon,photon] (39) -- (40);

\draw[color=Maroon,photon] (25) -- (41);
\draw[double]  (41) -- (42);
\draw[dash pattern=on 2pt off 1.5pt] (43) -- (42);

\end{tikzpicture}
    \end{gathered}  \\[5pt]
    &\Rightarrow \quad   y^d_{ij} \, q_i d^c_j h \quad \text{with} \quad y^d \sim \begin{bmatrix}
     0 & \epsilon^5   & \epsilon^5 \\
     \epsilon^5 & \epsilon^4 e^{i \theta} & \epsilon^4  \\
      0& 0& \epsilon^2 
      \end{bmatrix}.
\end{aligned}
\label{eq:DownChains}
\end{equation}
where again the black dots represent $D^c$, while the white dots $D$, and we have taken the Yukawa couplings $\hat{\lambda}_D \sim \epsilon$. The presence of a loop means that one of the links entering the loop must have a phase, and we chose it to be in entry $(2,2)$ of the low-energy Yukawa matrix -- $i.e.$ in the UV chain it comes from a link in the path between $d^c_2$ to $q_2 h$. A set of parameters which correctly match the down quark masses as well as the CKM matrix (at the TeV scale) is 
\begin{equation*}
\begin{aligned}
y^d_{12} \sim 6.0 \times 10^{-5}, &\quad y^d_{13} \sim 5.1 \times 10^{-5}, \\
y^d_{21} \sim 5.7 \times 10^{-5}, &\quad y^d_{22} \sim (2.5 \times 10^{-4}) \, e^{i 5.1},\\
y^d_{23} \sim 5.8 \times 10^{-4}, &\quad y^d_{33} \sim 1.4 \times 10^{-2},\\
\end{aligned}
\end{equation*}
which follows the parametric scaling in powers of $\epsilon$ given by Eq.\eqref{eq:DownChains}. 
\vspace{-3mm}

\section{Constraints}
\label{sec:Constraints}

We now turn to the experimental constraints on the class of models described above. In general, theories addressing the origin of flavor are subject to very stringent bounds from indirect searches, which typically push the new degrees of freedom associated with flavor dynamics well beyond the reach of direct detection. 
We show that this conclusion does not apply to the framework considered here. 
In particular, for values of $\epsilon \sim 0.1$, the scale of NP can naturally lie at the TeV scale, rendering these models potentially testable.
\\

\noindent {\bf EW Gauge Boson Couplings.}--- The presence of VL fermions with Yukawa couplings to the SM fields induces shifts in the couplings of the Higgs and the electroweak gauge bosons to SM fermions. Since we only allow hopping between $E^c/U^c/D^c$, only the couplings of left-handed fields are modified. If VL $L/Q$ states with Yukawa couplings to the light $e^c$ were also present, analogous shifts would arise for right-handed fields. We discuss these effects extensively in the SupM, and here we restrict ourselves to a few general observations. The relative shift that is diagonal in flavor space is generically of order $\sim \lambda^2 \frac{v^2}{M^2}$, while off-diagonal contributions are expected to be further suppressed by additional powers of $\epsilon$, provided that links between different chains occur far from the $\ell/q$'s sites. Precision measurements of the total and partial widths of the $Z$ boson constrain these corrections at the per-mil level~\cite{SLD:1996gjt,SLD:2000leq,SLD:2000ujp,SLD:2000jop,ALEPH:2005ab}. 
This implies a bound on VL masses of order $M_{E,U,D} \gtrsim (\lambda/0.2)\, {\rm TeV}$. These constraints are expected to improve by an order of magnitude at FCC-ee running at the $Z$ pole~\cite{Selvaggi:2025kmd}. Similar constraints can be derived by tests of lepton flavor universality in charged current mediated processes.
\\

\noindent {\bf Flavor.}--- Strong indirect constraints on these theories arise from flavor observables probing flavor violation in both the lepton and quark sectors. As discussed above, flavor off-diagonal effects are expected to be parametrically suppressed by powers of~$\epsilon$.
Indeed, the generation of an off-diagonal coupling requires a single vector-like fermion to couple simultaneously to two light generations. However, an additional and unavoidable source of flavor violation arises from the rotation between the light flavor and mass eigenstates. This misalignment induces flavor off-diagonal gauge boson couplings to fermions, even if the underlying interactions are flavor diagonal in the interaction basis. Detailed derivations and discussions of the relevant processes are provided in the SupM, and here we simply summarize their constraining power. 

In the quark sector, the most stringent constraints arise from neutral meson mixing, namely from $\Delta F = 2$ processes such as $K - \bar{K}\,, B_d - \bar{B}_d\,, B_s - \bar{B}_s\,, D - \bar{D}$. Let us focus on $K - \bar{K}$ mixing, as the remaining cases can be obtained straightforwardly by appropriate replacements of the flavor indices. For our class of models, since only left-handed quarks have Yukawa couplings to the new VL fermions, the only operator that is generated is the left--left operator $Q_1 = (\bar{s}\,\bar{\sigma}_\mu d)\, (\bar{s}\,\bar{\sigma}^\mu d)$. The corresponding Wilson coefficient $C_1$ receives contributions from tree-level $Z$ exchange, as well as from one-loop box diagrams involving the exchange of two heavy fermions. The dominant contribution is proportional to the left-diagonalizing unitary matrix $U_d$, and at leading order in $\epsilon$ we get 
\begin{equation*}
C_1 \simeq
\Bigl[\frac{2}{128\pi^2 M^2}+\frac{g_2^2 v^4}{8c_W^2 m_Z^2 M^4}\Bigr]
\Bigl[{\textstyle\sum_{i=1}^3} |\hat{\lambda}_i|^2 (U_d)_{i1}(U_d^*)_{i2}\Bigr]^2
\end{equation*}
where the factor of 2 in the first term comes from the inclusion of diagrams with $U^c$ and $D^c$ and we are assuming the mass of all VL fermions to be $\sim M$. First, we note that in the limit of exact flavor universality one finds $C_1 = 0$, so that the leading non-vanishing contributions arise only at higher order in $\epsilon$. If instead a $U(2)$ flavor symmetry acts on the first two generations, $C_1$ receives contributions proportional to two off-diagonal elements of $U_d$. More generally, taking $U_d = V_{\rm CKM}$ and forbidding accidental cancellations, we employ the constraints derived in Ref.~\cite{UTfit:2007eik} to obtain
\begin{equation}
    M_{U,D} \gtrsim\; \left[\frac{\lambda}{0.2},\, \left(\frac{\lambda}{0.2}\right)^2 \right]\, {\rm TeV} \;\;\;\; (Z \; {\rm tree}, hh \; {\rm box})\,.
\end{equation}
In the lepton sector, very stringent constraints arise from lepton flavor violating processes, such as $\mu \to e \gamma$, $\mu \to e$ conversion in nuclei, and $\ell \to 3\ell^{\prime}$ processes. 
Accommodating VL leptons at the TeV scale requires the mixing in the light charged-lepton sector to be strongly suppressed between the first two generations, $\mathcal{O}(10^{-6})$, while mixings involving the third generation can be as large as $\mathcal{O}(10^{-2})$. This is not in conflict with the observed mixing in the lepton sector, since it can arise entirely from the neutrino sector. In conclusion, for Yukawa couplings of order 0.2, the NP scale of these models can be as low as 1 TeV.\\

\noindent {\bf Direct Searches.}--- There are also direct searches for VL fermions at colliders. In $pp$ collisions, VL quarks (VLQs) can be produced in pairs via QCD interactions or singly through electroweak processes. For VLQs with $\Gamma/M \lesssim 5\%$, the limits are dominated by pair-production searches and are effectively independent of the width. Current analyses constrain VLQs that couple to third-generation SM quarks and decay primarily into $tZ, tH, bZ, bH$, to have masses above $\sim 1.5\, {\rm TeV}$, with the HL-LHC expected to strengthen these bounds by a few hundred GeV~\cite{CMS:2024bni}. To the best of our knowledge, dedicated searches for VLQs predominantly coupled to the first two SM quark generations have not yet been performed.\\
At the LHC, searches for VL leptons (VLLs) primarily target multilepton final states. At present, only VLL singlets under $SU(2)_L$ that couple predominantly to third-generation SM leptons are constrained, with current analyses setting a lower bound of roughly $200\,\rm{GeV}$ on their mass~\cite{CMS:2024bni}. Projections for the HL-LHC indicate significantly stronger sensitivity, with expected lower limits of approximately $600\,{\rm GeV}$, $650\, {\rm GeV}$, and $400\, {\rm GeV}$ for VLL singlets coupling mainly to first-, second-, and third-generation SM leptons, respectively~\cite{CMS:2024bni}.

While not yet competitive with flavor constraints, it is reasonable to expect that, in the near future, and with the implementation of search strategies tailored to the class of models discussed here, direct searches may surpass indirect ones. 
\section{Collider Signals}

The mechanism for explaining the origin of flavor presented in this work can be tested experimentally if the VL fermions are light enough to be produced at colliders and the structure of their Yukawa couplings to light fields can be probed. As we have seen, flavor constraints allow an ultra-low scale for the masses of the VL particles, so it is certainly possible that either the final phase of the LHC or future colliders, including Higgs/$Z$ factories, as well as 10 TeV scale muon colliders or a 100 TeV proton-proton collider, might be able to produce these states. Furthermore, these particles can decay \textit{only} via their Yukawa couplings to the light SM fields, with decays such as $E \to (\nu W_L,eZ_L,eh)$, $U \to (d W_L, u Z_L, u h)$, $D \to (u W_L, d Z_L, d h)$ with (2:1:1) branching ratios. Measuring these decays offers a possible probe of the size of these couplings. All of the VL particles can obviously be produced by pair production, via gauge interactions. 

The physics of VL L/Q decays differs importantly depending on whether the heavy fermions have generic masses differing by $O(1)$, or are nearly degenerate (within ${\cal O}(\epsilon)$).   
In the generic case, we have seen that the Yukawa couplings of the heavy fields are hierarchical. The VL states with the largest Yukawa couplings to the SM $\ell$'s and $q$'s, $\hat{\lambda}$, can also be singly produced via mixing. For these states, it is conceivable to measure their $\mathcal{O}(\epsilon)$ Yukawa coupling to light fields, both by measuring the rates for single production as well as measuring their decay widths:
\begin{equation}
    \Gamma = \bigg(\frac{|\hat{\lambda}|}{0.2}\bigg)^2  \bigg(\frac{M}{650\, {\rm GeV}}\bigg) \text{ GeV}\,, \text{ with } \hat{\lambda} \sim \epsilon.
\end{equation}
This is in principle within the reach of a clean environment at a muon collider. The VL fermions that are $p$-steps further down in the chain will have Yukawas  $\mathcal{O}( \epsilon^{p})$; if $p \leq 7$ the decays will still be prompt on the scale of the detector. Furthermore, for any $p>1$, the rate for single production will be minuscule, and the decay widths will be smaller than experimental resolution and so, unfortunately, neither of these methods will allow an easy test of the hierarchical pattern of (heavy-light) Yukawas. Nonetheless, there are two striking qualitative features that one could still look for: one of them simply reflects the fact that we need longer chains to explain smaller Yukawas. If the length of the chains is dictated by having the minimal size needed to generate correct size of the SM Yukawas, then the chain for $e$'s will be longer than that for $\mu$'s, which in turn will be longer than that for $\tau$'s. As a result, the number of VL states decaying to $e$'s will be bigger than those decaying to $\mu$'s and similar from $\mu$'s to $\tau$'s. It is also conceivable that the chains are longer than needed to generate flavor, for instance the $e^c,\mu^c,\tau^c$, could be connected not at the end of the chains but instead somewhere in the middle of the chains. 

The length of the chains is strictly constrained only by Landau poles of the gauge couplings, produced by the accelerated running induced by the new states. In particular for the model dicussed above, the number of VL pairs is $N_e =9$, $N_u =8$ and $N_d=19$, for $M \sim$ TeV in which case SU$(3)_c$ hits a Landau pole at $\sim 10^3$ TeV, of course SU$(2)_L$ is unaffected since the new matter are all SU$(2)_L$ singlets, and hypercharge alone would survive until $\sim 10^7$ TeV. There are obvious mechanisms that allow us to reduce the number of VL fermions. For instance, in models with 2 Higgs doublets, there can easily be an overall ``$1/\tan\beta \sim y_b/y_t \sim 1/50$'' suppression to all of the down and lepton Yukawas. In this case, in the lepton/down sectors there will be no chains for the third generation, and the remaining chains can have 2 fewer VL pairs, which takes the SU$(3)_c$ Landau pole to $10^{11}$ TeV. 

In this case, the lifetime of the particles with $p\geq 7$ distance down the chain, is long enough to lead to macroscopic decay lengths
\begin{equation}
    c \tau \sim 10^{2p-15} \text{cm} \times \left(\frac{0.1}{\epsilon} \right)^{2p} \times \frac{\text{TeV}}{M}\,.
\end{equation}
It would be interesting to look for other possible probes of Yukawa couplings for states with $p<7$. On the other hand, if the VL masses are nearly degenerate, then they can all have comparable Yukawas to the light fields. It is likely that the clean environment of the muon collider would allow the mass splittings $\delta M \sim \epsilon M \sim 100$ GeV to be measured, and their decay widths could also be measured. In addition, we have the interesting physics of oscillations between the nearly degenerate species, giving another handle on the mixings inherited from the chain structure. 

Apart from direct production, future colliders and in particular a future $Z$ factory producing $\sim 10^{12}$ $Z$'s can also indirectly probe the $\sim 10^{-2} \to 10^{-3}$ level of shifts to the $Z$ couplings induced by mixing with the heavy fields. We leave a detailed investigation of all these signals to future work. 
\section{Outlook}

We presented a simple four-dimensional mechanism to generate the SM flavor hierarchies where new VL fermions ``hop'' along discrete chains. Integrating out the heavy states generates effective Yukawas scaling as
$y_{ij}\sim \hat{\lambda} \epsilon^{n_{ij}}$, so the observed hierarchies can follow from moderately small parameters, $\hat{\lambda}, \epsilon\sim 0.1$-$0.2$. The site-link structure makes CP and flavor violation naturally mild: in the simplest lepton setup, disconnected open chains contain no rephasing-invariant phases and suppress flavor-changing effects, while in the quark sector a controlled loop structure can yield CKM-like CP violation. Existing indirect constraints (from precision electroweak data and flavor observables) are satisfied even for VL fermions near the TeV scale and we provide representative, phenomenologically viable, constructions. A simple chain model gives also rise to small neutrino masses (see Sec.\ref{sec:Neutrinos} of SupM), parametrically of order $y_e^2 v^2/M \sim 0.1 \, \mbox{eV} (\mbox{TeV}/M)$, and order unity mixing angles.

Notably, the model introduces solely vector-like fermions, absent additional gauge or scalar degrees of freedom; the absence of scalars in particular means that our construction is entirely agnostic about the hierarchy problem. Our framework could be a part of a more complete theory for the origin of the weak scale that is only slightly out of reach of the LHC, or could survive up to ultra-high energy scales with no additional matter. The ingredients used in our models are all standard and familiar in the vast literature of flavor model-building, so what is new? What allows us to comfortably have theories of flavor at the TeV scale rather than above the customary $\sim 10^2-10^4$ TeV? 

The phenomenological viability of the model rests on the soft flavor-symmetry breaking induced by hopping terms. In the heavy-mass diagonal basis, we already have the standard model Yukawa matrices, while interactions between heavy and light fields are suppressed by a universal factor $\epsilon$, alongside potential hierarchical suppressions for states that are further down the chain.

In the lepton sector we are free to imagine no link between the chains, which precludes flavor violation in the charged lepton sector and guarantees that all CP violation can be removed.
In the quark sector, restricting hopping to $(U^c, D^c)$ limits flavor violation to left-handed currents, and these are less constrained by FCNCs than the LR operators. For example, the LL Higgs box diagram for $K-\bar{K}$ mixing has a coefficient $\sim  \theta_c^2 \epsilon^4/ (128 \pi^2 M^2)$, allowing $M \sim 1$ TeV. Furthermore, a mild assumption about the coupling of the light quarks at the end of the chains -- that the couplings of the quarks at the end of the chains are democratically equal -- leads to an accidental $U(2)_q$ symmetry for all the leading couplings to heavy modes, mitigating even further this constraint, such that $M$ is bounded primarily by direct LHC searches and electroweak precision data. We can compare this with e.g. the gluino box diagram in supersymmetric theories, where $\epsilon^4 \to g_3^4$ and the most dangerous LR operators are generated, imposing strong constraints on the squark mass matrix. 

It is also obviously natural to compare our models with theories where the Yukawa hierarchy is generated from the geography of wavefunction overlap in flat or warped extra dimensions; in the warped case this is in some cases holographically dual to popular theories of partial compositeness~\cite{Csaki:2008zd,Glioti:2024hye,Agashe:2025tge}. The strongest constraints on these models come from electric dipole moments, as well as tree-level exchange of KK modes giving $K \bar{K}$ mixing at tree-level. These contributions typically necessitate a flavor scale in excess of $\sim 20$~TeV~\cite{Csaki:2008zd,Blanke:2008zb}. Crucially, our framework restricts lattice sites solely to the fermion sector, without additional light fields that have ``local" interactions on the chain. The absence of discretized gauge dynamics precludes the existence of heavy vector resonances (KK analogs), thereby eliminating the severe tree-level FCNC constraints intrinsic to standard deconstructed flavor architectures.

In this vein, we note the importance of simply declaring that the hopping parameters $\mu \sim \epsilon M$ softly break the flavor symmetry, without invoking new light fields near $M$ or $\mu$. Of course there are many simple field theoretic mechanisms for generating the small ratio $\epsilon \sim \mu/M \sim 0.1-0.2$ in the deep UV, and indeed it is not even unreasonable to think this modest small parameter already emerges in the low-energy theory beneath the Planck/string scales. But simply having soft breaking ``by hand" coming from the deeper UV contrasts with   conventional Froggatt-Nielsen phenomenology~\cite{Froggatt:1978nt, Leurer:1992wg}, where we would associate $\mu$ with the vev of light flavon scalar fields $\phi$. In this case, tree-level light flavon exchange might be problematic. In Sec.\ref{sec:TeVScalars} of SupM we examine this question in our set-up and find, remarkably, they are in fact {\it not} problematic for flavon and VL fermion masses as low as the TeV scale.

In conclusion, we have shown that the origin of flavor can occur at energies probed today at the LHC, and also at possible future colliders probing the 10 TeV scale. The framework's parsimony, characterized by simple particle content and sparse interactions, permits decisive experimental verification of its essential mechanisms if the new particles are within reach. While the presented examples are illustrative, we also believe that interesting new insights into the pattern of flavor mixing are to be found by a more systematic exploration of the ideas we have set forward here, which we intend to return to in future work.

\let\oldaddcontentsline\addcontentsline
\renewcommand{\addcontentsline}[3]{}

\section*{Acknowledgements}
The work of N.A.H. is supported by the DOE (Grant No. DE-SC0009988), the Simons Collaboration on Celestial Holography, the European Union (ERC, UNIVERSE PLUS, 101118787), and the Carl B. Feinberg cross-disciplinary program in innovation at the IAS. The work of C.F. is supported by FCT/Portugal (Grant No. 2023.01221.BD). The work of L.J.H. was supported by the NSF grant PHY-2515115 and the Office of High Energy Physics of the U.S. Department of
Energy under contract DE-AC02-05CH11231. The work of C.A.M. is supported by the U.S. Department of Energy (DE-SC0009988) and the Sivian Fund. Views and opinions expressed are those of the author(s) only and do not necessarily reflect those of the European Union or the European Research Council Executive Agency. Neither the European Union nor the granting authority can be held responsible for them.

\newpage

\bibliographystyle{apsrev4-1.bst}
\bibliography{Refs.bib}

\let\addcontentsline\oldaddcontentsline

\clearpage

\onecolumngrid
\begin{center}
  \textbf{\large Supplementary Material for New theories for the origin of flavor at the TeV scale}\\[.2cm]
  \vspace{0.05in}
  {Nima Arkani-Hamed, Carolina Figueiredo, Lawrence J. Hall, Claudio Andrea Manzari}
\end{center}

\twocolumngrid

\setcounter{equation}{0}
\setcounter{figure}{0}
\setcounter{table}{0}
\setcounter{section}{0}
\setcounter{page}{1}
\makeatletter
\renewcommand{\theequation}{S\arabic{equation}}
\renewcommand{\thefigure}{S\arabic{figure}}
\renewcommand{\thetable}{S\arabic{table}}

\setcounter{secnumdepth}{2}
\renewcommand{\thesection}{\Roman{section}}
\renewcommand{\thesubsection}{\thesection.\alph{subsection}}

\onecolumngrid

\startcontents[sections]

\section{Deriving low-energy constraints}

For simplicity, let us start by discussing what happens in the lepton sector. In a general chain model where there are links between chains, at low-energies we will generate off-diagonal couplings to the $Z$ and the $W$ bosons. Let us derive what this correction looks like. In the heavy mass basis, we have the following interactions:
\begin{equation}
  \mathcal{L} \supset M_I E_I E^c_I + \lambda^E_{i I} \ell_i E^c_I  h + y_{ij} \ell_i e^c_j h  
\end{equation}

After integrating out the $E/E^c$, we generate the following operator
\begin{equation}
\begin{gathered}
    \begin{tikzpicture}[line width=0.7,scale=0.3,baseline={([yshift=0.0ex]current bounding box.center)},
  fermion/.style={
    postaction={
      decorate,
      decoration={
        markings,
        mark=at position 0.6 with {
          \arrow{Latex[length=5pt,width=5pt]}
        }
      }
    }
  },
  scalar/.style={
    dashed,
    postaction={
      decorate,
      decoration={
        markings,
        mark=at position 0.6 with {
          \arrow{Latex[length=5pt,width=5pt]}
        }
      }
    }
  }
]

\coordinate (v0) at (0,4);
\coordinate (v1) at (2,2);
\coordinate (v2) at (0,0);
\coordinate (v5) at (6,2);
\coordinate (v8) at (8,4);
\coordinate (v9) at (8,0);

\coordinate (Ec) at (4,2);
\coordinate (li) at ($(v0)!0.5!(v1)$);
\coordinate (h) at ($(v2)!0.5!(v1)$);

\coordinate (lib) at ($(v5)!0.5!(v8)$);
\coordinate (hb) at ($(v5)!0.5!(v9)$);

\draw[fermion] (v1) -- (v0);
\draw[scalar]  (v1) -- (v2);
\draw[fermion] (v1) -- (v5);

\node[color=black,scale=0.9,xshift=0,yshift=9] at (Ec) {$E^c_I$};

\node[color=black,scale=0.9,xshift=5,yshift=9] at (li) {$\ell_{j}$};
\node[color=black,scale=0.9,xshift=5,yshift=-9] at (h) {$h^\dagger$};

\node[color=black,scale=0.9,xshift=-5,yshift=9] at (lib) {$\ell_{i}$};
\node[color=black,scale=0.9,xshift=-5,yshift=-9] at (hb) {$h$};

\draw[fermion] (v8) -- (v5);
\draw[scalar]  (v9) -- (v5);

\end{tikzpicture}
\end{gathered} \quad  + \quad \begin{gathered}
    \begin{tikzpicture}[line width=0.7,scale=0.29,baseline={([yshift=0.0ex]current bounding box.center)},
  fermion/.style={
    postaction={
      decorate,
      decoration={
        markings,
        mark=at position 0.7 with {
          \arrow{Latex[length=4.5pt,width=4.5pt]}
        }
      }
    }
  },
  antifermion/.style={
    postaction={
      decorate,
      decoration={
        markings,
        mark=at position 0.6 with {
          \arrow{Latex[length=4.5pt,width=4.5pt, reversed]}
        }
      }
    }
  },
  scalar/.style={
    dashed,
    postaction={
      decorate,
      decoration={
        markings,
        mark=at position 0.6 with {
          \arrow{Latex[length=4.5pt,width=4.5pt]}
        }
      }
    }
  },
  antiscalar/.style={
    dashed,
    postaction={
      decorate,
      decoration={
        markings,
        mark=at position 0.5 with {
          \arrow{Latex[length=4.5pt,width=4.5pt, reversed]}
        }
      }
    }
  },
  photon/.style={
    decorate,
    decoration={snake, amplitude=1.2pt, segment length=4pt},
  }
]

\coordinate (v0)  at (0,4);
\coordinate (v1)  at (2,2);
\coordinate (v2)  at (0,0);
\coordinate (v10) at (4,2);
\coordinate (v11) at (6,2);
\coordinate (v12) at (4,0);
\coordinate (v13) at (8,2);
\coordinate (v14) at (10,2);
\coordinate (v15) at (8,4);
\coordinate (v16) at (8,0);

\coordinate (Ec) at ($(v10)!0.5!(v1)$);
\node[color=black,scale=0.7,xshift=-2,yshift=9] at (Ec) {$E^c_I$};

\coordinate (E) at ($(v10)!0.5!(v11)$);
\node[color=black,scale=0.7,xshift=2,yshift=9] at (E) {$E^c_I$};

\draw[fermion]     (v1)  -- (v0);
\draw[scalar]      (v1)  -- (v2);
\draw[fermion]     (v1) -- (v10);

\draw[fermion]     (v10) -- (v11);
\draw[photon]      (v12) -- (v10);

\coordinate (B) at ($(v12)!0.5!(v10)$);
\node[color=black,scale=0.8,xshift=9,yshift=-7] at (B) {$B^\mu$};

\draw[antifermion] (v11) -- (v15);
\draw[antiscalar]  (v11) -- (v16);

\coordinate (li) at ($(v0)!0.5!(v1)$);
\coordinate (h) at ($(v2)!0.5!(v1)$);

\coordinate (lib) at ($(v15)!0.5!(v11)$);
\coordinate (hb) at ($(v16)!0.5!(v11)$);

\node[color=black,scale=0.9,xshift=5,yshift=9] at (li) {$\ell_{j}$};
\node[color=black,scale=0.9,xshift=3,yshift=-8] at (h) {$h^\dagger$};

\node[color=black,scale=0.9,xshift=-3,yshift=9] at (lib) {$\ell_{i}$};
\node[color=black,scale=0.9,xshift=-3,yshift=-8] at (hb) {$h$};

\end{tikzpicture}
\end{gathered} + \quad \begin{gathered}
    \begin{tikzpicture}[line width=0.7,scale=0.29,baseline={([yshift=0.0ex]current bounding box.center)},
  fermion/.style={
    postaction={
      decorate,
      decoration={
        markings,
        mark=at position 0.7 with {
          \arrow{Latex[length=4.5pt,width=4.5pt]}
        }
      }
    }
  },
  antifermion/.style={
    postaction={
      decorate,
      decoration={
        markings,
        mark=at position 0.6 with {
          \arrow{Latex[length=4.5pt,width=4.5pt, reversed]}
        }
      }
    }
  },
  scalar/.style={
    dashed,
    postaction={
      decorate,
      decoration={
        markings,
        mark=at position 0.6 with {
          \arrow{Latex[length=4.5pt,width=4.5pt]}
        }
      }
    }
  },
  antiscalar/.style={
    dashed,
    postaction={
      decorate,
      decoration={
        markings,
        mark=at position 0.5 with {
          \arrow{Latex[length=4.5pt,width=4.5pt, reversed]}
        }
      }
    }
  },
  photon/.style={
    decorate,
    decoration={snake, amplitude=1.2pt, segment length=4pt},
  }
]

\coordinate (v0)  at (0,4);
\coordinate (v1)  at (2,2);
\coordinate (v2)  at (0,0);
\coordinate (v10) at (4,2);
\coordinate (v11) at (6,2);
\coordinate (v12) at (6,0);
\coordinate (v13) at (8,2);
\coordinate (v14) at (10,2);
\coordinate (v15) at (12,4);
\coordinate (v16) at (12,0);

\coordinate (Ec) at ($(v10)!0.5!(v1)$);
\node[color=black,scale=0.7,xshift=-2,yshift=9] at (Ec) {$E^c_I$};

\coordinate (E) at ($(v10)!0.5!(v11)$);
\node[color=black,scale=0.7,xshift=2,yshift=9] at (E) {$E_I$};

\coordinate (Eb) at ($(v13)!0.5!(v11)$);
\node[color=black,scale=0.7,xshift=-1,yshift=9] at (Eb) {$E_I$};

\coordinate (Ecb) at ($(v13)!0.5!(v14)$);
\node[color=black,scale=0.7,xshift=3,yshift=9] at (Ecb) {$E^c_I$};

\draw[fermion]     (v1)  -- (v0);
\draw[scalar]      (v1)  -- (v2);
\draw[fermion]     (v1) -- (v10);

\node[cross, minimum size=2pt,scale=0.8] at (v10) {};
\node[cross, minimum size=2pt,scale=0.8] at (v13) {};

\draw[fermion]     (v11) -- (v10);
\draw[photon]      (v12) -- (v11);

\coordinate (B) at ($(v11)!0.5!(v12)$);
\node[color=black,scale=0.8,xshift=12,yshift=-7] at (B) {$B^\mu$};

\draw[antifermion] (v11) -- (v13);
\draw[antifermion] (v14) -- (v13);
\draw[antifermion] (v14) -- (v15);
\draw[antiscalar]  (v14) -- (v16);

\coordinate (li) at ($(v0)!0.5!(v1)$);
\coordinate (h) at ($(v2)!0.5!(v1)$);

\coordinate (lib) at ($(v15)!0.5!(v14)$);
\coordinate (hb) at ($(v16)!0.5!(v14)$);

\node[color=black,scale=0.9,xshift=5,yshift=9] at (li) {$\ell_{j}$};
\node[color=black,scale=0.9,xshift=5,yshift=-9] at (h) {$h^\dagger$};

\node[color=black,scale=0.9,xshift=-3,yshift=9] at (lib) {$\ell_{i}$};
\node[color=black,scale=0.9,xshift=-5,yshift=-9] at (hb) {$h$};

\end{tikzpicture}
\end{gathered}\quad =   (\ell_j h)^\dagger \bar{\sigma}^\mu D_\mu (\ell_i h) \; \frac{\lambda_{Ij}^\dagger  \lambda_{iI} }{M_{I}^2},
\label{eq:WfRen}
\end{equation}
where we are summing over $I$, corresponding to all VL fields that have Yukawas to $\ell_i$ and $\ell_j$ in the heavy mass basis. After spontaneous symmetry breaking, $h \to v$ (with $v=174$ GeV), this operator gives the following shift:
\begin{equation}
\mathcal{L} \supset \left[\delta_{ji} + \frac{\lambda_{Ij}^\dagger \lambda_{iI} v^2 }{M_{I}^2} \right] \bar{e}_j \bar{\sigma}^\mu D^{\text{\tiny EM}}_\mu e_i  - \frac{g_2}{c_W}(T_3 - Q s_W^2) \left[\delta_{j,i} + \frac{\lambda_{Ij}^\dagger \lambda_{iI} v^2 }{M_{I}^2} R\right] \bar{e}_j \bar{\sigma}^\mu Z_\mu e_i \quad \text{ with } R = \frac{-Q s_W^2}{T_3-Q s_W^2},
\end{equation}
where the $W^\mu$ couplings remain unaffected since the $D_\mu$ in \eqref{eq:WfRen} only contains $B_\mu$. Now, to canonically normalize the kinetic term, we appropriately rescale the $e$-fields so that at leading order in $v^2/M^2$, we obtain the following $Z^\mu$ coupling
\begin{equation}  
\mathcal{L}_Z \supset -\frac{g_2}{c_W}(T_3 - Q s_W^2) \left[\delta_{ji} + (R-1) \frac{\lambda_{Ij}^\dagger \lambda_{iI}  v^2 }{M_{I}^2} \right]\bar{e}_j \bar{\sigma}^\mu Z_\mu e_i+ \mathcal{O}(v^4/M^4),
\label{eq:Z-coup1}
\end{equation}
where $(R-1) = -T_3/(T_3 - Q s_W^2)$. In addition, after spontaneous symmetry breaking, we want to further diagonalize the light-light mass matrix $y_{i \alpha} v  e_i e^c_\alpha$, which at leading order in $v^2/M^2$ is still given by the Yukawa matrix, $y_{i\alpha}$. This is achieved by rotating $e_i \to (U^y_e)_{i,j} e_j$, and similar for $e^c$, and thus,  after diagonalizing the Yukawa/mass matrix, we obtain the final couplings to $Z^\mu$
\begin{equation}
\begin{aligned}
    \mathcal{L}_Z \supset  &\left[-\frac{g_2}{c_W}(T_3 - Q s_W^2) \delta_{kl} + (\delta G_Z^e)_{kl}\right]\bar{e}_k \bar{\sigma}^\mu Z_\mu e_l + \mathcal{O}(v^4/M^4)\\
    &\quad \text{ with }\quad  (\delta G_Z^e)_{kl} = \frac{g_2 T_3}{ c_W}v^2  (U_e^y)^\dagger_{kj}\frac{\lambda_{Ij}^\dagger \lambda_{iI} }{M_{I}^2} (U_e^y)_{il}
\end{aligned}
\label{eq:Z-coup2}
\end{equation}

Note that if the link between the different chains is placed somewhere in the middle, way from the Yukawa couplings to $\ell_i$, then the off-diagonal terms in $\lambda^\dagger \lambda/M^2$ will be suppressed by many powers of $\epsilon$. Therefore, the leading flavour-violating effect comes from the diagonal part of $\lambda^\dagger \lambda/M^2$ together with rotation angles in $U^y_e$.

The rescaling of $e_i/\bar{e}_i$ also induces a shift in the $W^\mu$ coupling, which at leading order in $v^2/M^2$ becomes 
\begin{equation}
\begin{aligned}
    \mathcal{L}_W \supset &  \left[\frac{g_2}{\sqrt{2}}(U_e^y)^\dagger_{ki}(U_\nu)_{il} + (\delta G_{W^-}^\ell)_{kl}  \right] \bar{e}_k \bar{\sigma}^\mu W^-_\mu \nu_l + \text{ h.c. } \\
    &\, \, \text{ with }\quad  (\delta G_{W^-}^\ell)_{kl} = - \frac{g_2 \,v^2}{2\sqrt{2}} (U_e^y)^\dagger_{kj}\frac{\lambda_{Ij}^\dagger \lambda_{iI} }{M_{I}^2}  (U_\nu)_{il}
\end{aligned}
\end{equation}
where $U_\nu$ is the rotation on the neutrinos, $\nu_j$. Finally, due to the original rescaling, we have that the low-energy mass matrix also receives $v^2/M^2$ corrections of the form
\begin{equation}
\begin{aligned}
   \mathcal{L}_{\text{Yuk.}} &\supset  v  e_i e^c_\alpha  
    \left[y_{i \alpha } - \frac{v^2}{2} y_{  j \alpha }\frac{\lambda_{Ij}^\dagger \lambda_{iI} }{M_{I}^2} + \mathcal{O}(v^4/M^4)  \right]  \\
    &\to m_i e_i e^c_i - \frac{v^3}{2}   (U_{e^c}^y)_{\alpha\beta} y_{j \alpha }\frac{\lambda_{Ij}^\dagger \lambda_{iI} }{M_{I}^2} (U_{e}^y)_{ik}  e_k e^c_{\beta} + \mathcal{O}(v^4/M^4) 
\end{aligned}
\end{equation}
where in the second line we show the result after rotating $e$ and $e^c$. In particular, we see explicitly that there are off-diagonal corrections to the Yukawa matrix suppressed by higher powers of $v^2/M^2$. Now, if we revert some $v$ back to the higgses, we see that these corrections are reflecting the presence of off-diagonal higgs couplings, concretely, at leading order in $v^2/M^2$, we have
\begin{equation}
    \mathcal{L}_\text{higgs} \supset \frac{y_{i \alpha }}{\sqrt{2}}  e_i  e^c_{\alpha} h  - \frac{3v^2}{4} y_{j \alpha }(U_{e^c}^y)_{\alpha\beta} \frac{\lambda_{Ij}^\dagger \lambda_{iI} }{M_{I}^2} (U_{e}^y)_{ik}   e_k  e^c_{\beta}h.
\end{equation}
Just like the off-diagonal $Z$ couplings, these interactions will also mediate flavor-violating processes. Nonetheless, the effects are much smaller as they are suppressed by the actual Yukawa matrix.

All of the results above, regarding the shift in the coupling of leptons to the $Z$ and $W$ bosons extend trivially to the quark sector. For a general chain model for the quark-sector, in the heavy mass basis we have the following interactions
\begin{equation}
  \mathcal{L} \supset M^u_I U_I U^c_I + M^d_J D_J D^c_J +  + \lambda^u_{iI}  q_i U^c_I  h^\dagger +  \lambda^d_{i J} q_i D^c_J h + y^u_{i \alpha} q_i u^c_\alpha h^\dagger  + y^u_{i \beta} q_i d^c_\beta h,
\end{equation}
so that after integrating out the VL fermions, $U/U^c$ and $D/D^c$, we generate, respectively 
\begin{equation}
     (q_j h^\dagger)^\dagger \bar{\sigma}^\mu D_\mu (q_i h^\dagger)  \frac{ \lambda_{Ij}^{u\,\dagger} \lambda^u_{iI} }{M^{u\, 2}_{I}}, \quad \text{ and } \quad (q_j h)^\dagger \bar{\sigma}^\mu D_\mu (q_i h)  \frac{\lambda^{d \,\dagger}_{Jj} \lambda^d_{iJ}}{M^{d\, 2}_{J}}.
     \label{eq:OperatorQuark}
\end{equation}

Note however that one can build chain models which only generate the flavor diagonal version of the operators above by ensuring that there is no path in the chains connecting $q_i$ to $q_j$ involving \textit{exclusively} VL fermions -- one example of such a model is the one given in the main text where $q_i$ and $q_j$ are connected to each other by paths that always include a massless $d^c$. But for the sake of generality let us stick with \eqref{eq:OperatorQuark}, where just like in the lepton sector, after performing a rotation+rescaling, we obtain the $Z$-couplings given in \eqref{eq:Z-coup1}, with the appropriate $T_3/Q$'s ($Q_u = 2/3$, $Q_d=-1/3$, and $T_3^u=1/2$, $T_3^d=-1/2$). 
Therefore, after we diagonalize both low-energy Yukawas/mass matrices, $u_i  \to (U^y_u)_{ij} u_j$ and $d_i  \to (U^y_d)_{ij} d_j$, at leading order in $v^2/M^2$, we obtain the analogous  shift of \eqref{eq:Z-coup2} for the quarks: 
\begin{equation}
(\delta G_Z^u)_{k,l} = \frac{g_2 T_3^u}{c_W}v^2  (U_u^y)^\dagger_{ki}\frac{ \lambda_{Ij}^{u\,\dagger} \lambda^u_{iI} }{M^{u\, 2}_{I}} (U_u^y)_{jl},  \quad \quad (\delta G_Z^d)_{k,l} =\frac{g_2 T_3^d}{ c_W}v^2  (U_d^y)^\dagger_{ki}\frac{\lambda^{d \,\dagger}_{Jj} \lambda^d_{iJ}}{M^{d\, 2}_{J}}(U_d^y)_{jl},
\end{equation}
and similarly for the $W$ couplings we obtain
\begin{equation}
\begin{aligned}
    \mathcal{L}_W \supset&  \left[\frac{g_2}{\sqrt{2}}(U^y_d)^{\dagger}_{ki} (U^y_u)_{il}  + (\delta G_{W^-}^q)_{kl}\right] \bar{d}_k \bar{\sigma}^\mu W^-_\mu u_l + \text{h.c.}\\
    & \, \text{ with }\quad (\delta G_{W^-}^q)_{kl} =  - \frac{g_2v^2}{2\sqrt{2}} (U^y_d)^{\dagger}_{ki}\left[ \frac{ \lambda_{Ij}^{u\,\dagger} \lambda^u_{iI} }{M^{u\, 2}_{I}} + \frac{\lambda^{d \,\dagger}_{Jj} \lambda^d_{iJ}}{M^{d\, 2}_{J}} \right] (U^y_u)_{jl}.
\end{aligned}
\end{equation}

Finally just like we saw for the leptons, the higgs will also have off-diagonal interactions with the $u$'s and the $d$'s which read
\begin{equation}
    \mathcal{L}_\text{higgs} \supset - \frac{3v^2}{2}\left[ (U_{u}^y)_{ki}  \frac{ \lambda_{Ij}^{u\,\dagger} \lambda^u_{iI} }{M^{u\, 2}_{I}} y^u_{j \alpha}(U_{u^c}^y)_{\alpha\rho} u_k u^c_{\rho} + (U_{d}^y)_{ki} \frac{\lambda^{d \,\dagger}_{Jj} \lambda^d_{iJ}}{M^{d\, 2}_{J}}y^d_{j \alpha} (U_{d^c}^y)_{\beta\rho} d_k d^c_{\rho}\right] \, h ,
\end{equation}
which are once again further suppressed by a power of $y^u/y^d$ as compared to the off-diagonal $Z$-couplings.

\subsection{Constraints from flavor}

Let us now discuss the leading flavor violating processes which are allowed by the presence of the VL fermions. For the derivations bellow we will assume the most general chain model (with arbitrary mixings and links between chains), however, as explained in the main text, many of these constraints can be evaded in specific models. 
\\ 

\paragraph{Lepton Sector: $\ell_i \to \ell_f \gamma$ ---} If there are links between the heavy $E/E^c$ between different chains, then at low-energies we have $E^c$'s with Yukawas to different light generations, and thus we can have $\ell_i \to \ell_f \gamma$. This process is described by the following dipole operators and respective Wilson coefficient:
\begin{equation}     \mathcal{L}_{\text{eff}} = a^{if} \,e^c_f \sigma^{[\mu} \bar{\sigma}^{\nu]} e_i  F_{\mu\nu} + b^{if}  \, \bar{e}_f \bar{\sigma}^{[\mu} \sigma^{\nu]} \bar{e}^c_i  F_{\mu\nu},
\label{eq:effL_MuEGam}
\end{equation}
which are generated, at leading order, by the following diagrams: 
\begin{equation}
\begin{gathered}
        \begin{tikzpicture}[line width=0.7,scale=0.32,baseline={([yshift=0.0ex]current bounding box.center)},
   fermion/.style={
    postaction={
      decorate,
      decoration={
        markings,
        mark=at position 0.6 with {
          \arrow{Latex[length=4.5pt,width=4.5pt]}
        }
      }
    }
  },
  antifermion/.style={
    postaction={
      decorate,
      decoration={
        markings,
        mark=at position 0.6 with {
          \arrow{Latex[length=4.5pt,width=4.5pt, reversed]}
        }
      }
    }
  },
  scalar/.style={
    dashed,
    postaction={
      decorate,
      decoration={
        markings,
        mark=at position 0.65 with {
          \arrow{Latex[length=4.5pt,width=4.3pt]}
        }
      }
    }
  },
  antiscalar/.style={
    dashed,
    postaction={
      decorate,
      decoration={
        markings,
        mark=at position 0.5 with {
          \arrow{Latex[length=4.5pt,width=4.5pt, reversed]}
        }
      }
    }
  },
  photon/.style={
    decorate,
    decoration={snake, amplitude=1.2pt, segment length=4pt},
  }
]

\coordinate (v0) at (0,4);
\coordinate (v1) at (2,4);
\coordinate (v2) at (4,2);
\coordinate (v3) at (6,4);
\coordinate (v4) at (4,0);
\coordinate (v5) at (8,4);

\coordinate (l2) at ($(v0)!0.5!(v1)$);
\node[color=black,scale=0.7,xshift=3,yshift=9] at (l2) {$\ell_i$};

\coordinate (l1) at ($(v3)!0.5!(v5)$);
\node[color=black,scale=0.7,xshift=3,yshift=9] at (l1) {$\ell_f$};

\coordinate (h) at ($(v3)!0.5!(v1)$);
\node[color=black,scale=0.7,xshift=3,yshift=9] at (h) {$h$};

\coordinate (Ec) at ($(v1)!0.5!(v2)$);
\node[color=black,scale=0.7,xshift=-7,yshift=-7] at (Ec) {$E^c_I$};

\coordinate (Ecb) at ($(v3)!0.5!(v2)$);
\node[color=black,scale=0.7,xshift=8,yshift=-7] at (Ecb) {$\overline{E}^c_I$};

\coordinate (gam) at ($(v4)!0.5!(v2)$);
\node[color=black,scale=0.7,xshift=11,yshift=-7] at (gam) {$B_\mu$};

\draw[fermion]     (v0) -- (v1);
\draw[antifermion] (v1) -- (v2);
\draw[fermion]     (v3) -- (v2);
\draw[scalar]      (v3) -- (v1);
\draw[photon]      (v4) -- (v2);
\draw[fermion]     (v3) -- (v5);

\fill (v1) circle (1.6pt);
\fill (v2) circle (1.6pt);
\fill (v3) circle (1.6pt);

\end{tikzpicture}
    \end{gathered}\quad ,\quad \begin{gathered}
        \begin{tikzpicture}[line width=0.7,scale=0.32,baseline={([yshift=0.0ex]current bounding box.center)},
   fermion/.style={
    postaction={
      decorate,
      decoration={
        markings,
        mark=at position 0.7 with {
          \arrow{Latex[length=4.5pt,width=4.5pt]}
        }
      }
    }
  },
  antifermion/.style={
    postaction={
      decorate,
      decoration={
        markings,
        mark=at position 0.6 with {
          \arrow{Latex[length=4.5pt,width=4.5pt, reversed]}
        }
      }
    }
  },
  scalar/.style={
    dashed,
    postaction={
      decorate,
      decoration={
        markings,
        mark=at position 0.65 with {
          \arrow{Latex[length=4.5pt,width=4.3pt]}
        }
      }
    }
  },
  antiscalar/.style={
    dashed,
    postaction={
      decorate,
      decoration={
        markings,
        mark=at position 0.5 with {
          \arrow{Latex[length=4.5pt,width=4.5pt, reversed]}
        }
      }
    }
  },
  photon/.style={
    decorate,
    decoration={snake, amplitude=1.2pt, segment length=4pt},
  }
]

\coordinate (v0) at (0,4);
\coordinate (v1) at (2,4);
\coordinate (v2) at (4,2);
\coordinate (v3) at (6,4);
\coordinate (v4) at (4,0);
\coordinate (v5) at (8,4);

\coordinate (l2) at ($(v0)!0.5!(v1)$);
\node[color=black,scale=0.7,xshift=3,yshift=9] at (l2) {$\ell_i$};

\coordinate (l1) at ($(v3)!0.5!(v5)$);
\node[color=black,scale=0.7,xshift=3,yshift=9] at (l1) {$\ell_f$};

\coordinate (h) at ($(v3)!0.5!(v1)$);
\node[color=black,scale=0.7,xshift=3,yshift=9] at (h) {$h$};

\coordinate (Ec) at ($(v1)!0.5!(v2)$);
\node[color=black,scale=0.7,xshift=-16,yshift=0] at (Ec) {$E^c_I$};
\node[color=black,scale=0.7,xshift=2,yshift=-15] at (Ec) {$E_I$};

\coordinate (Ecb) at ($(v3)!0.5!(v2)$);

\coordinate (gam) at ($(v4)!0.5!(v2)$);
\node[color=black,scale=0.7,xshift=11,yshift=-7] at (gam) {$B_\mu$};

\node at (Ec) {
  \tikz[rotate=45]{
    \draw[line width=1.2pt, white] (-2pt,-2pt)--(2pt,2pt) (-2pt,2pt)--(2pt,-2pt);
    \draw[line width=0.6pt, black] (-2pt,-2pt)--(2pt,2pt) (-2pt,2pt)--(2pt,-2pt);
  }
};

\node at (Ecb) {
  \tikz[rotate=45]{
    \draw[line width=1.2pt, white] (-2pt,-2pt)--(2pt,2pt) (-2pt,2pt)--(2pt,-2pt);
    \draw[line width=0.6pt, black] (-2pt,-2pt)--(2pt,2pt) (-2pt,2pt)--(2pt,-2pt);
  }
};

\draw[fermion]     (v0) -- (v1);
\draw[antifermion] (v1) -- (Ec);
\draw[antifermion] (v2) -- (Ec);
\draw[fermion]     (v3) -- (Ecb);
\draw[fermion]     (v2) -- (Ecb);
\draw[scalar]      (v3) -- (v1);
\draw[photon]      (v4) -- (v2);
\draw[fermion]     (v3) -- (v5);

\fill (v1) circle (1.6pt);
\fill (v2) circle (1.6pt);
\fill (v3) circle (1.6pt);

\end{tikzpicture}
    \end{gathered}  \quad ,\quad \begin{gathered}
        \begin{tikzpicture}[line width=0.7,scale=0.32,baseline={([yshift=0.0ex]current bounding box.center)},
   fermion/.style={
    postaction={
      decorate,
      decoration={
        markings,
        mark=at position 0.6 with {
          \arrow{Latex[length=4.5pt,width=4.5pt]}
        }
      }
    }
  },
  antifermion/.style={
    postaction={
      decorate,
      decoration={
        markings,
        mark=at position 0.6 with {
          \arrow{Latex[length=4.5pt,width=4.5pt, reversed]}
        }
      }
    }
  },
  scalar/.style={
    dashed,
    postaction={
      decorate,
      decoration={
        markings,
        mark=at position 0.65 with {
          \arrow{Latex[length=4.5pt,width=4.3pt]}
        }
      }
    }
  },
  antiscalar/.style={
    dashed,
    postaction={
      decorate,
      decoration={
        markings,
        mark=at position 0.5 with {
          \arrow{Latex[length=4.5pt,width=4.5pt, reversed]}
        }
      }
    }
  },
  photon/.style={
    decorate,
    decoration={snake, amplitude=1.2pt, segment length=4pt},
  }
]

\coordinate (v0) at (0,4);
\coordinate (v1) at (2,4);
\coordinate (v2) at (4,2);
\coordinate (v3) at (6,4);
\coordinate (v4) at (4,0);
\coordinate (v5) at (8,4);

\coordinate (l2) at ($(v0)!0.5!(v1)$);
\node[color=black,scale=0.7,xshift=3,yshift=9] at (l2) {$\ell_i$};

\coordinate (l1) at ($(v3)!0.5!(v5)$);
\node[color=black,scale=0.7,xshift=3,yshift=9] at (l1) {$\ell_f$};

\coordinate (h) at ($(v3)!0.5!(v1)$);
\node[color=black,scale=0.7,xshift=3,yshift=9] at (h) {$E^c_I$};

\coordinate (Ec) at ($(v1)!0.5!(v2)$);
\node[color=black,scale=0.7,xshift=-7,yshift=-7] at (Ec) {$h$};

\coordinate (Ecb) at ($(v3)!0.5!(v2)$);

\coordinate (gam) at ($(v4)!0.5!(v2)$);
\node[color=black,scale=0.7,xshift=21,yshift=-7] at (gam) {$B_\mu/ W_\mu$};

\draw[fermion]     (v0) -- (v1);
\draw[antiscalar] (v1) -- (v2);
\draw[scalar]     (v3) -- (v2);
\draw[fermion]      (v3) -- (v1);
\draw[photon]      (v4) -- (v2);
\draw[fermion]     (v3) -- (v5);

\fill (v1) circle (1.6pt);
\fill (v2) circle (1.6pt);
\fill (v3) circle (1.6pt);

\end{tikzpicture}
    \end{gathered}.
\end{equation}
In addition to the diagrams above, there are also contributions from diagrams with a larger number of higgs insertions, but these are suppressed by powers of $v^2/M^2$. 
After rotating the fields appropriately to diagonalize the low-energy mass matrix, and working at leading order in $m_H/M$ we find the following Wilson coefficients:
\begin{equation}    
a^{i f} = (U^y_e)^\dagger_{f,j}\frac{ \lambda^\dagger_{I j} \lambda_{k I}}{M_I^2}(U^y_e)_{k,i} \times \frac{e m_f}{192 \pi^2}, \quad \quad b^{i f} =  (U^y_e)^\dagger_{f,j}\frac{ \lambda^\dagger_{I j} \lambda_{k I}}{M_I^2} (U^y_e)_{k,i}\times \frac{e m_i}{192 \pi^2} .  
\end{equation}
Finally, the branching ratio for $\ell_i \to \ell_f \gamma$ can be written directly in terms of these Wilson coefficients as follows
\begin{equation}
{\rm Br}[\ell_i \to \ell_f \gamma]=\frac{m_{\ell_i}^3}{4\pi \, \Gamma_{i}} \big(|a^{if} |^{2}+ |b^{if} |^{2}\big),
\label{Brmuegamma}
\end{equation}
and the experimental bounds (at 90\% C.L.) are~\cite{MEGII:2025gzr,BaBar:2009hkt}
\begin{equation}
\operatorname{Br}(\mu \rightarrow e \gamma)\leq 1.5 \times 10^{-13}, \quad 
\operatorname{Br}(\tau \rightarrow e \gamma)\leq 3.3 \times 10^{-8}, \quad 
\operatorname{Br}(\tau \rightarrow \mu \gamma)\leq 4.4 \times 10^{-8}.
\end{equation}
Looking at the most stringent case of $\mu \to e \gamma$, and assuming that the biggest Yukawa $\lambda \sim \epsilon$, and that there is an order $\epsilon$ rotation angle between the first and second generation in $U_e^y$, we obtain the following constraint: 
\begin{equation}
    \operatorname{Br}(\mu \rightarrow e \gamma) \sim \left(\frac{17\, \text{GeV}}{M_I} \right)^4 \times \epsilon^6 \leq 1.5 \times 10^{-13}, \quad \Rightarrow \quad M_I \gtrsim (1-3) \text{ TeV},
\end{equation}
where we took $\epsilon \in [0.1,0.2]$. Note, however, that as emphasized in the text, this is almost the worst possible scenario -- there are different ways of placing the links between the chains such that there are more powers of epsilon and therefore $M_I$ is allowed to be beneath 1 TeV.  
\\

\paragraph{Lepton Sector: $l \to 3l $ ---} The presence of off-diagonal couplings of the $Z$ at low energies also leads to $l \to 3l $ decays. Concretely, the leading contribution to these processes comes from:
\begin{equation}
\begin{gathered}
    \begin{tikzpicture}[line width=0.7,scale=0.55,baseline={([yshift=0.0ex]current bounding box.center)},
  fermion/.style={
    postaction={
      decorate,
      decoration={
        markings,
        mark=at position 0.6 with {
          \arrow{Latex[length=4.5pt,width=4.5pt]}
        }
      }
    }
  },
  antifermion/.style={
    postaction={
      decorate,
      decoration={
        markings,
        mark=at position 0.6 with {
          \arrow{Latex[length=4.5pt,width=4.5pt, reversed]}
        }
      }
    }
  },
  scalar/.style={
    dashed,
    postaction={
      decorate,
      decoration={
        markings,
        mark=at position 0.6 with {
          \arrow{Latex[length=4.5pt,width=4.5pt]}
        }
      }
    }
  },
  antiscalar/.style={
    dashed,
    postaction={
      decorate,
      decoration={
        markings,
        mark=at position 0.5 with {
          \arrow{Latex[length=4.5pt,width=4.5pt, reversed]}
        }
      }
    }
  },
  photon/.style={
    decorate,
    decoration={snake, amplitude=1.2pt, segment length=4pt},
  }
]

\coordinate (v0)  at (0,2);
\coordinate (v1)  at (2,2);
\coordinate (v8)  at (3.5,3);
\coordinate (v9)  at (3.5,1);
\coordinate (v10) at (4.8,1.8);
\coordinate (v11) at (4.8,0.6);

\draw[fermion] (v0) -- (v1);
\draw[fermion] (v1) -- (v8);
\draw[photon]  (v1) -- (v9);
\draw[fermion] (v9) -- (v10);
\draw[fermion] (v11) -- (v9);

\coordinate (li) at ($(v0)!0.5!(v1)$);
\node[color=black,scale=0.7,xshift=-11,yshift=5] at (li) {$e_i$};

\coordinate (lbj) at ($(v1)!0.5!(v8)$);
\node[color=black,scale=0.7,xshift=4,yshift=10] at (lbj) {$e_j$};

\coordinate (Z) at ($(v1)!0.5!(v9)$);
\node[color=black,scale=0.7,xshift=-4,yshift=-10] at (Z) {$Z_\mu$};

\coordinate (ebk) at ($(v9)!0.5!(v10)$);
\node[color=black,scale=0.7,xshift=4,yshift=10] at (ebk) {$e_k$};

\coordinate (ek) at ($(v9)!0.5!(v11)$);
\node[color=black,scale=0.7,xshift=10,yshift=4] at (ek) {$e_k$};

\end{tikzpicture}
\end{gathered} = \frac{G_Z^e  (\delta G_Z^e)_{ji}}{m_Z^2}  (\bar{e}_j \bar{\sigma}^\mu e_i)(\bar{e}_k \bar{\sigma}_\mu e_k) , \quad \quad \begin{gathered}
    \begin{tikzpicture}[line width=0.7,scale=0.55,baseline={([yshift=0.0ex]current bounding box.center)},
  fermion/.style={
    postaction={
      decorate,
      decoration={
        markings,
        mark=at position 0.6 with {
          \arrow{Latex[length=4.5pt,width=4.5pt]}
        }
      }
    }
  },
  antifermion/.style={
    postaction={
      decorate,
      decoration={
        markings,
        mark=at position 0.6 with {
          \arrow{Latex[length=4.5pt,width=4.5pt, reversed]}
        }
      }
    }
  },
  scalar/.style={
    dashed,
    postaction={
      decorate,
      decoration={
        markings,
        mark=at position 0.6 with {
          \arrow{Latex[length=4.5pt,width=4.5pt]}
        }
      }
    }
  },
  antiscalar/.style={
    dashed,
    postaction={
      decorate,
      decoration={
        markings,
        mark=at position 0.5 with {
          \arrow{Latex[length=4.5pt,width=4.5pt, reversed]}
        }
      }
    }
  },
  photon/.style={
    decorate,
    decoration={snake, amplitude=1.2pt, segment length=4pt},
  }
]

\coordinate (v0)  at (0,2);
\coordinate (v1)  at (2,2);
\coordinate (v8)  at (3.5,3);
\coordinate (v9)  at (3.5,1);
\coordinate (v10) at (4.8,1.8);
\coordinate (v11) at (4.8,0.6);

\draw[fermion] (v0) -- (v1);
\draw[fermion] (v1) -- (v8);
\draw[photon]  (v1) -- (v9);
\draw[fermion] (v9) -- (v10);
\draw[fermion] (v11) -- (v9);

\coordinate (li) at ($(v0)!0.5!(v1)$);
\node[color=black,scale=0.7,xshift=-11,yshift=5] at (li) {$e_i$};

\coordinate (lbj) at ($(v1)!0.5!(v8)$);
\node[color=black,scale=0.7,xshift=4,yshift=10] at (lbj) {$e_j$};

\coordinate (Z) at ($(v1)!0.5!(v9)$);
\node[color=black,scale=0.7,xshift=-4,yshift=-10] at (Z) {$Z_\mu$};

\coordinate (ebk) at ($(v9)!0.5!(v10)$);
\node[color=black,scale=0.7,xshift=4,yshift=10] at (ebk) {$e^c_k$};

\coordinate (ek) at ($(v9)!0.5!(v11)$);
\node[color=black,scale=0.7,xshift=10,yshift=4] at (ek) {$e^c_k$};

\end{tikzpicture}
\end{gathered}  = \frac{G_Z^{e^c}  (\delta G_Z^e)_{ji}}{m_Z^2} (\bar{e}_j \bar{\sigma}^\mu e_i)(\bar{e}^c_k \bar{\sigma}_\mu e^c_k).
\end{equation}
with $G_Z^{e} = -g_2 (T^e_3-s_W^2Q_e)/c_W$ and $G_Z^{e^c} = g_2s_W^2 Q_{e^c}/c_W$.

Let us focus on the decays involving only one flavour change, as exotic ones involving multiple flavour changes, such as $\tau^-\to e^-\mu^+e^-$, are further suppressed. The relevant branching ratios are 
\begin{equation}
\begin{aligned}
{\rm Br}(\mu\to 3e) = \frac{m_{\mu}^5}{1536\pi^3m_Z^4\Gamma_{\mu}}&\left[2|G_Z^e  (\delta G_Z^e)_{e\mu}|^2+|G_Z^{e^c}  (\delta G_Z^e)_{e\mu}|^2 \right] \,,\\
{\rm Br}(\tau\to e\mu\mu) = \frac{m_{\tau}^5}{1536\pi^3m_Z^4\Gamma_{\tau}}&\left[ \, |G_Z^e  (\delta G_Z^e)_{e\tau}|^2+|G_Z^{e^c}  (\delta G_Z^e)_{e\tau}|^2\right] \,,
\end{aligned}
\end{equation}
with $\Gamma_\mu, \, \Gamma_\tau$ being the muon and tau decay widths and $\xi$ the coupling of the $Z$ to leptons in the mass basis. The other processes can be read out with a trivial exchange of flavor indices. The corresponding experimental limits (at 90\% CL~\cite{HFLAV:2019otj,SINDRUM:1987nra,BaBar:2010axs,Hayasaka:2010np,LHCb:2014kws}) are given by
\begin{align}
\begin{split}
\operatorname{Br}(\mu \rightarrow e e e)&\leq 1.0 \times 10^{-12}\,,\\
\operatorname{Br}(\tau \rightarrow \mu \mu \mu) &\leq  1.1 \times 10^{-8}\,,\\
\operatorname{Br}(\tau \rightarrow e e e)&\leq 1.4 \times 10^{-8}\,, \\
\operatorname{Br}(\tau \rightarrow e \mu \mu) &\leq 1.6 \times 10^{-8}\,,\\
\operatorname{Br}(\tau \rightarrow \mu e e) &\leq 8.4 \times 10^{-9}\,.
\end{split}
\end{align}
So, once again, looking at the most string case of $\mu \to eee$, we find that the leading order effect for $\lambda \sim \epsilon$, and a rotation of $\epsilon$ betweeen first two generations, gives a bound on $M_I$:
\begin{equation}
    \operatorname{Br}(\mu \rightarrow e e e) \sim \left( \frac{117 \text{ GeV}}{M_I} \right)^4 \times \epsilon^6, \quad \Rightarrow \quad M_I \gtrsim (3-10) \text{ TeV},
\end{equation}
where we take $\epsilon \in [0.1,0.2]$.
\\

\paragraph{Lepton Sector: $e_i \to e_j$ conversion ---} Again, due to the presence of off-diagonal $Z$-couplings, we can have $\mu \to  e$ conversion in nuclei. In general, the relevant operators for this process are 
\begin{equation}
\mathcal{L}_{\text{eff}}=\sum_{q=u,d}\left(C_{qq}^{V\,LL}O_{qq}^{V\,LL}+C_{qq}^{V\,LR}O_{qq}^{V\,LR}\right)+(L\leftrightarrow R)+ \text{h.c.}\,,
\end{equation}

However, since in our models we only have $E/E^c$ coupling to $\ell_i$, we only have off-diagonal couplings of the $Z$ to $e_i$ (not $e^c_i$), and therefore we only generate the $LL$ and $LR$ operators which are
\begin{equation}
    O_{qq}^{V\,LL}=(\bar{e}_i \bar{\sigma}^\mu e_j)(\bar{q} \bar{\sigma}_\mu q)\,,\quad \quad 
O_{qq}^{V\,LR}=(\bar{e}_i \bar{\sigma}^\mu e_j)(q^c \bar{\sigma}_\mu \bar{q}^c)\,.
\end{equation}
The contribution to these operators  at leading order in $(v^2/M^2)$ comes from the tree-level diagrams with a $Z$ exchange, which yields
\begin{equation}
\begin{gathered}
    \begin{tikzpicture}[line width=0.7,scale=0.55,baseline={([yshift=0.0ex]current bounding box.center)},
  fermion/.style={
    postaction={
      decorate,
      decoration={
        markings,
        mark=at position 0.6 with {
          \arrow{Latex[length=4.5pt,width=4.5pt]}
        }
      }
    }
  },
  antifermion/.style={
    postaction={
      decorate,
      decoration={
        markings,
        mark=at position 0.5 with {
          \arrow{Latex[length=4.5pt,width=4.5pt, reversed]}
        }
      }
    }
  },
  scalar/.style={
    dashed,
    postaction={
      decorate,
      decoration={
        markings,
        mark=at position 0.6 with {
          \arrow{Latex[length=4.5pt,width=4.5pt]}
        }
      }
    }
  },
  antiscalar/.style={
    dashed,
    postaction={
      decorate,
      decoration={
        markings,
        mark=at position 0.5 with {
          \arrow{Latex[length=4.5pt,width=4.5pt, reversed]}
        }
      }
    }
  },
  photon/.style={
    decorate,
    decoration={snake, amplitude=1.2pt, segment length=4pt},
  }
]
\coordinate (v1) at (2,2);
\coordinate (v2) at (4,2);
\coordinate (v4) at (2,0);
\coordinate (v5) at (4,0);
\coordinate (v6) at (0,2);
\coordinate (v7) at (0,0);
\coordinate (v8) at (6,0);
\coordinate (v9) at (6,2);

\coordinate (qi) at ($(v1)!0.5!(v6)$);
\node[color=black,scale=0.8,xshift=-10,yshift=-7] at (qi) {$\bar{e}_i$};
\coordinate (Uci) at ($(v1)!0.5!(v2)$);
\node[color=black,scale=0.8,xshift=10,yshift=-8] at (Uci) {$e_j$};

\coordinate (qj) at ($(v7)!0.5!(v4)$);
\node[color=black,scale=0.8,xshift=-10,yshift=7] at (qj) {$\bar{q}/\bar{q}^c$};

\coordinate (Ucj) at ($(v4)!0.5!(v5)$);
\node[color=black,scale=0.8,xshift=10,yshift=7] at (Ucj) {$q/q^c$};

\draw[antifermion] (v1) -- (v2);
\draw[photon]  (v1) -- (v4);
\draw[antifermion]     (v6) -- (v1);
\draw[antifermion] (v7) -- (v4);
\draw[antifermion]     (v4) -- (v5);

\coordinate (h2) at ($(v1)!0.5!(v4)$);
\node[color=black,scale=0.7,xshift=-10,yshift=0] at (h2) {$Z^\mu$};

\end{tikzpicture}
\end{gathered} \, \, \Rightarrow \, \, C_{qq}^{V \, LL} = \frac{G_{Z}^{q} (\delta G^e_Z)_{ij}}{m_Z^2} , \quad C_{qq}^{V \, LR} = \frac{G_{Z}^{\bar{q}^c} (\delta G^e_Z)_{ij}}{m_Z^2},
\end{equation}
where $G_{Z}^{q}/G_Z^{\bar{q}^c}$ are the SM couplings of the quarks $u/\bar{u}^c$ and $d/\bar{d}^c$ to the $Z$-boson. The transition rate $\Gamma_{\mu\to e}^N\equiv \Gamma(\mu N\to eN)$ is given by (see e.g.~\cite{Cirigliano:2009bz,Crivellin:2014cta,Crivellin:2017rmk})
\begin{equation}
\Gamma_{\mu\to e}^N = 4 m_\mu^5 \,\Bigg \vert \sum_{q=u,d}\left(C_{qq}^{V\;LR}+C_{qq}^{V\;LL}\right)\left(f_{Vp}^{(q)}V_N^p\,
+\, f_{Vn}^{(q)}V_N^n\right)
\Bigg\vert^2\,,
\end{equation}
with $f_{V}^{(q)}$ being the nucleon vector form factors $f_{Vp}^{(u)}=2,\;f_{Vn}^{(u)}=1,\;f_{Vp}^{(d)}=1,\;f_{Vn}^{(d)}=2$; and $V_N^{p,n}$ are the overlap integrals. The general result for a nucleus N is
\begin{align}
    \Gamma_{\mu\to e}^N \simeq 5\times 10^{-5} \, {\rm GeV}^5\, (-0.5\, V_N^n + 0.04\, V_N^p)^2\, \Bigg \vert U_e^{y\,\dagger} \frac{\lambda_{I}^\dagger \lambda_{I} }{M_{I}^2}\, U_e^y \Bigg \vert^2\,.
\end{align}
For gold we use the numerical values~\cite{Kitano:2002mt}, $V_{\text{Au}}^p=0.0974\,, V_{\text{Au}}^n=0.146$, and after normalising by the capture rate~\cite{Suzuki:1987jf}, $\Gamma_{\text{Au}}^\text{capt}=8.7\times 10^{-18}\; \text{GeV}$, we can compare to the experimental 90\% C.L. limit~\cite{SINDRUMII:2006dvw}
\begin{equation}
\begin{aligned}
\frac{\Gamma_\text{Au}^\text{conv}}{\Gamma_\text{Au}^\text{capt}} &\simeq 3\times 10^{10} \times {\rm GeV}^4 \times \Bigg \vert U_e^{y\,\dagger} \frac{\lambda_{I}^\dagger \lambda_{I} }{M_{I}^2}\, U_e^y \Bigg \vert^2 < 7.0\times 10^{-13}\, \\
&\simeq \left( \frac{ 415 \text{ GeV}}{M_I} \right)^4  \times \epsilon^6 < 7.0\times 10^{-13} \quad \Rightarrow \quad  M_I \gtrsim (15-40) \text{ TeV}
\end{aligned}
\end{equation}
where we are again considering the most dramatic case where $U^\dagger_e \lambda^\dagger \lambda U_e \sim \epsilon^3$, and taking $\epsilon \sim 0.1-0.2$.
The near-future experiments Mu2e~\cite{Bernstein:2019fyh} and COMET~\cite{COMET:2018wbw} will measure muon-to-electron conversion in $^{27}$Al and are expected to significantly improve the current limits, reaching sensitivities of $\Gamma_\text{Al}^\text{conv}/\Gamma_\text{Al}^\text{capt} < 10^{-17}$, with $\Gamma_\text{Al}^\text{capt} = 2.76\times 10^{-34}$~GeV.\\

\paragraph{Lepton Sector: Tests of Lepton Flavor Universality ---}

\begin{table}[t!]
	\centering
	\begin{tabular}{l c c }
		\hline\hline
		Observable & Ref. & Measurement \\
		\hline 
		\\
		$R\left[\frac{K\rightarrow\mu\nu}{K\rightarrow e\nu}\right]\simeq|1+ \frac{\sqrt{2}}{g_2}( \delta G_{W^-}^\ell)_{2,2}- \frac{\sqrt{2}}{g_2}( \delta G_{W^-}^\ell)_{1,1}|$ &~\cite{Pich:2013lsa} &$0.9978 \pm 0.0020$ \\[0.2 cm]		
		$R\left[\frac{\pi\rightarrow\mu\nu}{\pi\rightarrow e\nu}\right]\simeq|1+ \frac{\sqrt{2}}{g_2}( \delta G_{W^-}^\ell)_{2,2}- \frac{\sqrt{2}}{g_2}( \delta G_{W^-}^\ell)_{1,1}|$&~\cite{PiENu:2015seu,ParticleDataGroup:2018ovx} & $1.0010 \pm 0.0009$ \\[0.2 cm]		
		$R\left[\frac{\tau\rightarrow\mu\nu\bar{\nu}}{\tau\rightarrow e\nu\bar{\nu}}\right]\simeq|1+ \frac{\sqrt{2}}{g_2}( \delta G_{W^-}^\ell)_{2,2}- \frac{\sqrt{2}}{g_2}( \delta G_{W^-}^\ell)_{1,1}|$&~\cite{HFLAV:2019otj,ParticleDataGroup:2018ovx} & $1.0018 \pm 0.0014$ \\[0.2 cm]		
		$R\left[\frac{K\rightarrow\pi\mu\bar{\nu}}{K\rightarrow\pi e\bar{\nu}}\right]\simeq|1+ \frac{\sqrt{2}}{g_2}( \delta G_{W^-}^\ell)_{2,2}- \frac{\sqrt{2}}{g_2}( \delta G_{W^-}^\ell)_{1,1}|$&~\cite{Pich:2013lsa} & $1.0010 \pm 0.0025$ \\[0.2 cm]		
		$R\left[\frac{W\rightarrow\mu\bar{\nu}}{W\rightarrow e\bar{\nu}}\right]\simeq|1+ \frac{\sqrt{2}}{g_2}( \delta G_{W^-}^\ell)_{2,2}- \frac{\sqrt{2}}{g_2}( \delta G_{W^-}^\ell)_{1,1}|$&~\cite{Pich:2013lsa,ALEPH:2013dgf} & $0.996 \pm 0.010$ \\	[0.2 cm]
		$R\left[\frac{\tau\rightarrow e\nu\bar{\nu}}{\mu\rightarrow e\bar{\nu}\nu}\right]\simeq|1+ \frac{\sqrt{2}}{g_2}( \delta G_{W^-}^\ell)_{3,3}- \frac{\sqrt{2}}{g_2}( \delta G_{W^-}^\ell)_{2,2}|$&~\cite{HFLAV:2019otj,ParticleDataGroup:2018ovx} & $1.0010 \pm 0.0014$ \\[0.2 cm]			
		$R\left[\frac{\tau\rightarrow \pi\nu}{\pi\rightarrow \mu\bar{\nu}}\right]\simeq|1+ \frac{\sqrt{2}}{g_2}( \delta G_{W^-}^\ell)_{3,3}- \frac{\sqrt{2}}{g_2}( \delta G_{W^-}^\ell)_{2,2}|$&~\cite{HFLAV:2019otj}& $0.9961 \pm 0.0027$ \\[0.2 cm]			
		$R\left[\frac{\tau\rightarrow K\nu}{K\rightarrow \mu\bar{\nu}}\right]\simeq|1+ \frac{\sqrt{2}}{g_2}( \delta G_{W^-}^\ell)_{3,3}- \frac{\sqrt{2}}{g_2}( \delta G_{W^-}^\ell)_{2,2}|$&~\cite{HFLAV:2019otj} & $0.9860 \pm 0.0070$ \\[0.2 cm]
		$R\left[\frac{W\rightarrow \tau\bar{\nu}}{W\rightarrow \mu\bar{\nu}}\right]\simeq|1+ \frac{\sqrt{2}}{g_2}( \delta G_{W^-}^\ell)_{3,3}- \frac{\sqrt{2}}{g_2}( \delta G_{W^-}^\ell)_{2,2}|$&~\cite{Pich:2013lsa,ALEPH:2013dgf,ATLAS:2020wvq} & \begin{tabular}{@{}c @{}} $1.034 \pm 0.013|_{\text{LEP}}$\\  $0.092\pm 0.013|_{\text{ATLAS}}$ \end{tabular} \\[0.2 cm]			
		$R\left[\frac{\tau\rightarrow \mu\nu\bar{\nu}}{\mu\rightarrow e\nu\bar{\nu}}\right]\simeq|1+ \frac{\sqrt{2}}{g_2}( \delta G_{W^-}^\ell)_{3,3}- \frac{\sqrt{2}}{g_2}( \delta G_{W^-}^\ell)_{1,1}|$&~\cite{HFLAV:2019otj,ParticleDataGroup:2018ovx} & $1.0029 \pm 0.0014$ \\[0.2 cm]			
		$R\left[\frac{W\rightarrow \tau\bar{\nu}}{W\rightarrow e\bar{\nu}}\right]\simeq|1+ \frac{\sqrt{2}}{g_2}( \delta G_{W^-}^\ell)_{3,3}- \frac{\sqrt{2}}{g_2}( \delta G_{W^-}^\ell)_{1,1}|$&~\cite{Pich:2013lsa,ALEPH:2013dgf} & $1.031 \pm 0.013$\\[0.2 cm]
		$R\left[\frac{B\rightarrow D^{(*)}\mu\nu}{B\rightarrow D^{(*)}e\nu}\right]\simeq|1+ \frac{\sqrt{2}}{g_2}( \delta G_{W^-}^\ell)_{2,2}- \frac{\sqrt{2}}{g_2}( \delta G_{W^-}^\ell)_{1,1}|$&~\cite{Jung:2018lfu} & $0.989 \pm 0.012$\\[0.2 cm]
		\hline\hline
	\end{tabular}	\caption{Ratios testing LFU together with their dependence on the modified $W\ell\nu$ couplings and the corresponding experimental values. Deviations from unity measures LFU violation.}\label{ObsLFU}
\end{table}

Violations of lepton flavor universality (LFU) in charged-current interactions can be probed through ratios of $W$, kaon, pion, and tau decay rates involving different charged leptons in the final state. Such ratios provide particularly stringent tests of LFU-violating effects, as they benefit from both reduced experimental and theoretical uncertainties. The observables considered in this work are listed in Table~\ref{ObsLFU}. For each observable $Y$, we define $R(Y)$ as the ratio of the corresponding amplitude in the presence of VLLs to that in the SM. By construction, $R(Y)=1$ in the limit of vanishing mixing between the SM leptons and the VLLs. As can be readily inferred from Table~\ref{ObsLFU}, the strongest experimental limits are at the per-mille level and give $ M \gtrsim 4\, \lambda$ TeV. In the near future, the PIONEER experiment~\cite{PIONEER:2022alm} is expected to improve on $R\left[\frac{\pi\rightarrow\mu\nu}{\pi\rightarrow e\nu}\right]$ by one order of magnitude.
\\

\paragraph{Quark Sector: Meson mixing---} Finally in the quark sector, the strongest constraints come from meson mixing, in particular processes with $\Delta F=2$, which are given by the following $4$-fermion operators: 
\begin{equation}
\mathcal{H}_{\text{eff}}^{\Delta F=2} = \sum_{i=1}^{5}C_iQ_i + \sum_{i=1}^{3}\tilde{C}_i\tilde{Q}_i , \, \, \,  \text{ with } \quad \begin{aligned}
    Q_1= (\bar{s}_{\alpha}\bar{\sigma}_{\mu}d_{\alpha})&(\bar{s}_{\beta}\bar{\sigma}^{\mu}d_{\beta}),\, \, \,  Q_2 = (s^{c}_{\alpha}d_{\alpha})(s^{c}_{\beta}d_{\beta}), \, \, \,  
Q_3 = (s^{c}_{\alpha}d_{\beta})(s^{c}_{\beta}d_{\alpha}), \\
 &Q_4 = (s^{c}_{\alpha}d_{\alpha})(\bar{s}_{\beta}\bar{d}^{c}_{\beta})\,,\, \, \,Q_5 = (s^{c}_{\alpha}d_{\beta})(\bar{s}_{\beta}\bar{d}^{c}_{\alpha})\,,
\end{aligned}
\label{eq:HDeltaF2}
\end{equation}
where we have given the explicit definition of $Q_i$ for the case of $K-\bar{K}$ mixing; $\tilde Q_i$ are operators obtained from $Q_i$ flipping the chirality of both fermion bilinears. For $B_d, B_s$ and $D$ mixing, the effective Hamiltonian is obtained by exchanging the quark fields appropriately. Once again, since our models only have VL $U/U^c$ and $D/D^c$, we will only generate the LL operator $Q_1$. The Wilson coefficient of $Q_1$ receives comparable contributions from the tree-level exchange of a $Z$ boson and the one-loop box diagrams with the exchange of two VL fermions.

The leading contribution comes from the VL $U/U^c$ and $D/D^c$ with Yukawas to the $q$'s -- that are at the beginning of the chains, whose effective Yukawas are denoted as $\lambda^{U/D}_i \sim \epsilon$, and which generate the operators in Eq.~\ref{eq:HDeltaF2} diagonal in flavor space. Upon rotating to the quark mass basis and at leading order in the mass splitting between the different VL fermions, $M^2 = M_U^2 + \delta M_U^2 = M_D^2 + \delta M_D^2$, we obtain
\begin{equation}
\begin{split}
      C_1^{\text{box}} = & \frac{1}{128 \pi^2 M^2} \sum_{a\in \{D,U\}} \left[\left|\lambda_1^a \right|^2 (U_d^y)_{11} (U_d^y)^\dagger_{12} +\left|\lambda_2^a \right|^2 (U_d^y)_{12} (U_d^y)^\dagger_{22} \right]^2  \\
      & \, \, + \frac{1}{128 \pi^2 M^2}   \sum_{a\in \{D,U\}} \sum_{i=1,2}2 \left|\lambda^a_{i}\right|^2 \left|\lambda^a_{3}\right|^2    (U_d^y)_{1i} (U_d^y)^\dagger_{i2}
      (U_d^y)_{13} (U_d^y)^\dagger_{32} + \left|\lambda^a_{3}\right|^4 \left[(U_d^y)_{13} (U_d^y)^\dagger_{32}\right]^2\,,
\end{split}
\end{equation}
and
\begin{equation}
\begin{split}
C_1^{Z} = &\frac{g_2^2 v^4}{8 c_W^2 m_Z^2 M^4}\left[\left|\lambda_1^D \right|^2 (U_d^y)_{11} (U_d^y)^\dagger_{21} +\left|\lambda_2^D \right|^2 (U_d^y)_{12} (U_d^y)^\dagger_{22} \right]^2  \\
&+\frac{g_2^2 v^4}{8 c_W^2 m_Z^2 M^4} \sum_{i=1,2}2 \left|\lambda^D_{i}\right|^2 \left|\lambda^D_{3}\right|^2    (U_d^y)_{1i} (U_d^y)^\dagger_{i2}
      (U_d^y)_{13} (U_d^y)^\dagger_{32} + \left|\lambda^D_{3}\right|^4 \left[(U_d^y)_{13} (U_d^y)^\dagger_{32}\right]^2\,.
\end{split}
\end{equation}

First of all, note that at this order, CP violation is only introduced through the rotation matrices. Moreover, from both of these contributions it is clear that if we further assume an effective $U(2)$ symmetry between the first two generations up to this order in $\epsilon$, so that $\lambda_1^{U/D}=\lambda_2^{U/D}$, then using the unitarity of $U$, the leading contribution turns into something involving also the third generation, which is further suppressed, are therefore making the model extra safe. More generally, taking $U^y_d = V_{\rm CKM}$ and forbidding accidental cancellations, we employ the constraints derived in Ref.~\cite{UTfit:2007eik} to obtain
\begin{equation}
    M_{U,D} \gtrsim\; \left[\frac{\lambda}{0.2},\, \left(\frac{\lambda}{0.2}\right)^2 \right]\, {\rm TeV} \;\;\;\; (Z \; {\rm tree}, hh \; {\rm box})\,.
\end{equation}
Additional constraints on VL leptons and VL quarks come from the flavour-changing neutral current transition $b \to s \gamma$~\cite{Paul:2016urs} and from data on Parity violation electron scattering off proton and nuclei~\cite{Crivellin:2021bkd}. In our scenario, these constraints are not competitive and allow NP below the TeV scale for $\lambda\sim 0.2$.

\section{Neutrino Masses}
\label{sec:Neutrinos}
Key questions of flavor physics involve neutrinos, including: what sets the scale of neutrino masses and why are the leptonic mixing angles order unity when those in the quark sector are small? In chain models there are simple answers to these questions. Here we consider theories having an Abelian flavor group, $G_F$, and no light right-handed neutrinos, so the observed neutrinos have Majorana masses with normal ordering. The mixing angles are large because $G_F$ does not distinguish between the three lepton doublets, $\ell_i$, and the neutrinos are very light because the chains are long and must involve a link that changes lepton number by two units. We illustrate this by extending the model shown in Fig. \ref{fig:LeptonChains} to include neutrino masses.

The leptonic sector of this model has $G_F = U(1)^3$, with one $U(1)$ for each generation. All leptonic states of generation $i$ can have charge under $U(1)_i$, but have zero charge under $U(1)_j, \, j \neq i$. The heavy states $E_{iq}$ are conveniently labelled by i, for those in the $\ell_i$ chain, and by $q$, which is the charge under $U(1)_i$. We choose $\ell_i$ to have charge 6 under $U(1)_i$ and $e^c_{1,2,3}$ to have charges (0,2,4) under $U(1)_{1,2,3}$. We choose the flavor-violating soft masses in the $\ell_i$ chains to all have charge $-1$ under $U(1)_i$. The $\ell_{1,2,3}$ chains then have (5,3,1) links hopping along the $E_{i,q}$, which have charges $q = (1, 2, 3, 4, 5; \,3, 4, 5; \, 5)$ for $i=(1;2;3)$, leading to the key result of (\ref{eq:EYukawas}).

\begin{figure}[t]
    \centering
    \includegraphics[width=\linewidth]{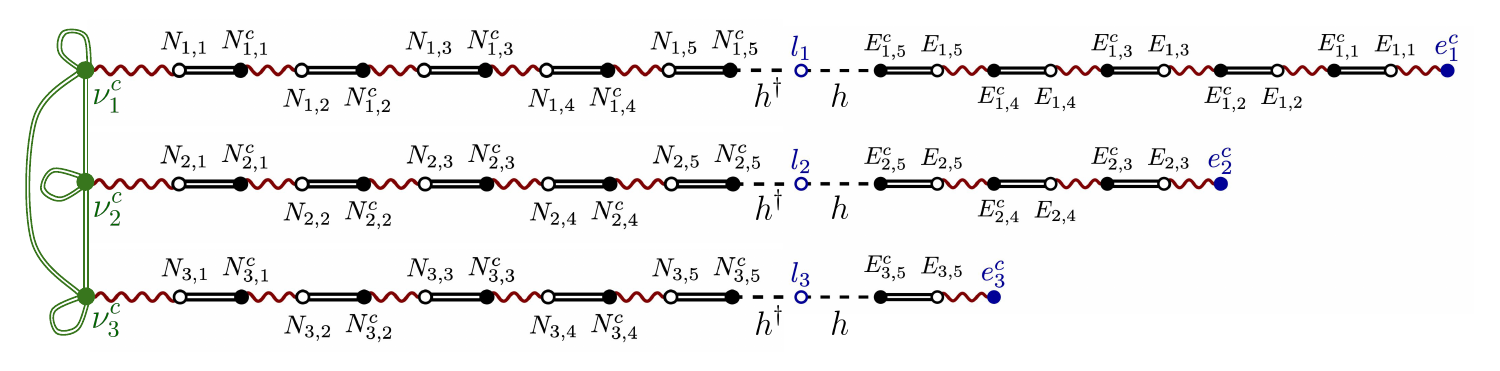}
    \caption{Chains for neutrinos. In addition to the chains presented in the main text (fig. \ref{fig:LeptonChains}), we have $3$ new chains of SM singlet states, $N_{i,q}$. At the end of these new chains, we have singlets that are also neutral under the flavor symmetry, $\nu^c_{i}$, which mix via a Majorana mass term $M_{ij} \nu^c_{i}\nu^c_{j}$, represented in green.}
    \label{fig:neutrinos}
\end{figure}

The neutrino sector is very similar to the lepton sector. There are a set of heavy VLF Dirac states that are SM singlets, $N_{i,q}$.  The key difference is that we choose $q=1-5$ for each $i$. We also introduce analogues of the $e^c_i$ states, $N^c_{i,0}$, which we call $\nu^c_i$.   The $i$ neutrino chains begin with the interactions $\lambda_i \ell_i h N^c_{i,5}$ and, since the flavor-violating soft masses again all have charge -1, after 5 links the chain ends in $\nu^c_{i}$ (see figure \ref{fig:neutrinos}). If this were the end of the story, there would be mixing between $\nu_i$ and $N_{i,q}$ states, but the neutrinos would remain massless because lepton number is unbroken. Crucially, the states $\nu^c_{i}$ have no gauge or flavor charges, and hence nothing forbids the Majorana mass term $M_{ij} \nu^c_{i} \nu^c_{j}$.  This is the sole violation of lepton flavor in the model and it allows any two neutrino chains to be pasted together to generate $(\ell_i h^\dagger)(\ell_j h^\dagger)$ operators for neutrino masses. The $G_F$-invariant Lagrangian describing lepton flavor is 
\begin{equation}
\begin{aligned}
  {\cal L}_{G_F}^{E,N} = \left(\lambda_i^E \, \ell_i E^c_{i,5} h + M_{iq}^E \, E_{i,q} E^c_{i,q}\right) + \left( \lambda_i^N \, \ell_i N^c_{i,5} h + M_{iq}^N \, N_{i,q} N^c_{i,q} + M_{ij} \, \nu^c_{i} \nu^c_{j} \right) + h.c. 
\end{aligned}
\label{eq:LENGFinv}
\end{equation}
where sums over $i,j$ and $q$ are understood.  The only differences between the $E$ and $N$ sectors are the number of terms in the sums on $q$ and the Majorana mass term $M_{ij}$.
The $G_F$-violating masses that provide nearest neighbor hopping along the chains are
\begin{equation}
\begin{aligned}
  {\cal L}_{G_F \, \text{viol}}^{E,N} \; = \; \mu_{iq}^E \, E_{i,q} E^c_{i,q-1} + \mu_i^E \, E_{i,q_-} e^c_i+ \mu_{iq}^N \, N_{i,q} N^c_{i,q-1} + h.c. 
\end{aligned}
\label{eq:LENGFviol}
\end{equation}
where $q_-$ is the lowest $q$ for $E_{iq}$. While the number of $N^c$ and $N$ states are equal, gauge anomalies force one more $E^c$ than $E$ per generation, and the additional states are the $e^c$ states of the SM. Taking all the $\lambda$ parameters equal, all the $M$ parameters equal and all the $\mu$ parameters equal, gives $y_e \sim \lambda \epsilon^5 \sim \epsilon^6$. Since the neutrino chains are of equal length, and have the same length as the chain for $y_e$, we predict order unity leptonic mixing angles and neutrino masses proportional to $y_e^2$
\begin{equation}
\begin{aligned}
  \theta_{ij} \; \approx \; 1, \hspace {0.5in} m_{\nu_i} \; \approx \; y_e^2 \; \frac{v^2}{M} \; \approx \; 0.1 \, \mbox{eV} \; \frac{\mbox{TeV}}{M}.
\end{aligned}
\label{eq:nu masses}
\end{equation}
These estimates are highly approximate as there could be a pileup of order unity couplings.

The similarity between the neutrino and charged lepton sectors is enhanced if the number of $E_{iq}$ states is the same for each $i$; indeed, the $E_{iq}$ and $N_{iq}$ states could have the same set of charges. In this case the hierarchy of charged lepton masses results because $e^c_i$ have different charges under $U(1)_i$. Some of the $E_{iq}$ are then ``spectator" VLF; they do not appear in the chains linking $\ell_i$ to $e^c_i$. If we imposed lepton number and added light $\nu^c_i$ states with the same charges as $e^c_i$, the forms of the charged lepton and neutrino sectors would become identical; the neutrinos would be Dirac with masses of order $m_e, m_\mu$ and $m_\tau$ and there would be no lepton flavor mixing. The striking differences between neutrinos and charged leptons arise simply because such $\nu^c_i$ states are absent, and the gauge and flavor symmetries allow $M_{ij} \, \nu^c_i \nu^c_j$.  

The doublet neutrino state $\nu_i$ and the singlet state $N_i$ closest to it in the chain mix with angle $\theta_i$, modifying the $W \bar{\nu}_i e_i $ coupling by $\cos \theta_i$ and the $Z \bar{\nu}_i \nu_i $ coupling by $\cos^2 \theta_i$. The mixing angle is $\theta_i \approx (\lambda_i v/M_i) \approx (\lambda_i/6)(\mbox{TeV}/M_i)$, leading to the expectation of non-universalities close to $10^{-3}$, the current experimental limit, for $\lambda_i \sim 0.2$ and $M_i \sim$ TeV. 

\section{Spontaneous Breaking of the Flavor Symmetry at the TeV scale}
\label{sec:TeVScalars}
In this letter we have softly broken the flavor group, $G_F$, via a set of fermion masses, $\mu^{U,D,E}_{IJ}$, that are smaller than the masses, $M^{U,D,E}_I$, of the VLF. How can these soft parameters arise? They can be as small as 100-200 GeV if the VLF are close to their experimental lower bounds at the TeV scale.  Remarkably, they can be generated by a generalization of the Froggatt-Nielsen mechanism. We introduce a scalar field that has Yukawa couplings $x$ and acquires a vev $\Phi = V + \phi$ spontaneously breaking $G_F$ so that 
\begin{equation}
\begin{aligned}
  \mu = x V, \hspace{0.5in} \epsilon = \frac{xV}{M},
\end{aligned}
\label{eq:mu=xV}
\end{equation}
where we have suppressed all flavor indices. Indeed, depending on $G_F$ and on quantum number assignments, $\Phi$ could be a set of fields. 

The scalar $\phi$ will have couplings to quarks and leptons, and in the down sector we expect powerful constraints from neutral kaon mixing. Integrating out the VLF in the chain with $n_{ij}$ links that generates the mass term $d_i d^c_j$ induces an operator involving $\Phi^{n_{ij}}$. Expanding about the vacuum, leads to a Yukawa interaction
\begin{equation}
\begin{aligned}
  {\cal L}_\phi \; = \; z^\prime_{ij} \; d^\prime_i d^{c \, \prime}_j \,\phi, \hspace{0.5in} z^\prime_{ij} \; = \; n_{ij} \,y^d_{ij} \, \frac{v}{V}.
\end{aligned}
\label{eq:zij}
\end{equation}
On rotating to the mass eigenstate basis, we have interactions $z_{ij} d_i d^c_j \,\phi$, which are not diagonal because $n_{ij}$ depends on $i,j$. While there are many possible models, assuming no small parameters in the potential for $\Phi$, we generically expect tree-level $\phi$ exchange to generate the operator $\mathcal{O}_4$ of (\ref{eq:HDeltaF2}), which has the most stringent constraint from the $K_L$-$K_S$ mass difference 
\begin{equation}
\begin{aligned}
  |C_4| = \frac{|z_{12} z^*_{21}|}{M_\phi^2} \; \leq \; \frac{4 \times 10^{-9}}{\mbox{TeV}^2} \hspace{0.5in} \Longrightarrow \hspace{0.5in} M_\phi V \geq (0.4 \, \mbox{TeV})^2
\end{aligned}
\label{eq:C4fromphi}
\end{equation}
at 95\% confidence level. To obtain the bound on $M_\phi V$ we took $z_{12,21}$ to be $\sqrt{y_d y_s} \, v/V$; there is no $n_{ij}$ factor because $n_{12,21} = n_{22}+1$. With no small parameters in the scalar potential, the $M_\phi$ and $V$ are comparable, so the 0.4 TeV bound applies approximately to both.  The key parameter setting flavor hierarchies, $\epsilon$ of (\ref{eq:mu=xV}), can be small from the Yukawa $x$ being less than unity and/or from the vev $V$ being less than the VLF masses $M$. For example, $\epsilon=0.2$ arises from ($x=0.2, \, V=M=1$ TeV) or from ($x=1, V=0.4$ TeV, $M=2$ TeV).

CP violation in kaon mixing implies an important constraint on theories with spontaneous $G_F$ breaking. If $C_4$ has comparable real and imaginary components, the limit from $\epsilon_K$ forces $V$ above 4 TeV, and to avoid $x < \epsilon$ also requires $M > 4$ TeV, suppressing collider signals. However, in the simplest set of theories, those where the diagram describing the chains in the quark sector has a single loop that spans both up and down sectors, there is no CP violation in tree-level $\phi$ exchange. The single loop implies that there is a single irremovable physical phase, and by rephasing quark fields it can reside on any link in the loop. Since the loop spans both up and down sectors, the phase can be rotated into the up sector so that $C_4$ is real.

If $G_F$ is discrete there are no Goldstone bosons. It may be that relevant operators in the potential possess an accidental symmetry, in which case some scalars are lighter than $V$, but we take the cutoff scale to be small enough that they evade experimental bounds. If $G_F$ contains a continuous symmetry, we must allow soft breaking in operators of dimension 2 to make the PGBs, $a$, sufficiently heavy. In the flavor basis, all Weyl fermions with non-zero Abelian charge will have a vector current coupled to $\partial^\mu a/V$. Rotating to the mass eigenstate basis then generates flavor-changing interactions, because $G_F$ acts differently on the three generations, for example $d_i \sigma_\mu \bar{d}_j \, \partial^\mu a/V$. If the scale $V$ is low we must require $m_a > m_{K,D,B}$ to avoid too-large rates for rare meson decays; avoiding this we still have to contend with FCNC's mediated by these light pseudos. There is a rich phenomenology of constraints and signals associated with these particles depending on their mass and decay constants, whose investigation we leave to future work. For the purposes of our letter, it is enough to note that these constraints and signals are optional, as we can assume that the PGBs are either not present to begin with or made massive enough by soft breaking to be at around the same mass as the generic $\phi$ fields in our discussion above.

\end{document}